\documentclass[useAMS,usenatbib]{mn2e}
\usepackage{graphicx}

\newcommand{\lsim}{\raisebox{-0.13cm}{~\shortstack{$<$ \\[-0.07cm] $\sim$}}~}
\newcommand{\gsim}{\raisebox{-0.13cm}{~\shortstack{$>$ \\[-0.07cm] $\sim$}}~}
\newcommand{\ffm}{$\rm 4.5 \, \mu m \,$}

\title[The stellar mass function at $3\leq z <5$]{The stellar mass function of the most massive galaxies at $3\leq z <5$ in the UKIDSS Ultra Deep Survey}
\author[K. I. Caputi et al.]{K. I. Caputi$^{1}$\thanks{E-mail:
kic@roe.ac.uk}, M. Cirasuolo$^{1,2}$,
J. S. Dunlop$^{1}$, R. J. McLure$^{1}$, D. Farrah$^{3}$
\newauthor and O. Almaini$^{4}$\\
$^{1}$SUPA\thanks{Scottish Universities Physics Alliance}, Institute for Astronomy, The University of Edinburgh, Royal Observatory, Edinburgh, EH9 3HJ\\
$^{2}$UK Astronomy Technology Center, Royal
Observatory, Blackford Hill, Edinburgh, EH9 3HJ\\
$^{3}$ Astronomy Centre, Department of Physics and Astronomy, University of Sussex, Brighton BN1 9QH\\
$^{4}$School of Physics and Astronomy, University of Nottingham, University Park, Nottingham NG7 2RD\\
}
\begin{document}

\date{Accepted 2010... Received 2010 ...; in original form 2010 ...}

\pagerange{\pageref{firstpage}--\pageref{lastpage}} \pubyear{2010}

\maketitle

\label{firstpage}

\begin{abstract}
We have analysed a sample of 1292 \ffm-selected galaxies at $z\geq 3$, over 0.6 deg$^2$ of the UKIRT Infrared Deep Survey (UKIDSS) Ultra Deep Survey (UDS). Using photometry from the $U$ band through \ffm, we have obtained photometric redshifts and derived stellar masses for our sources. Only two of our galaxies potentially lie at $z>5$. We have studied the galaxy stellar mass function  at $3\leq z< 5$, based on the 1213 galaxies in our catalogue with $[4.5] \leq 24.0$. We find that: i) the number density of $M \gsim 10^{11} \, \rm M_\odot$ galaxies increased by a factor $> 10$ between $z=5$ and 3, indicating that the assembly rate of these galaxies proceeded $> 20$ times faster at these redshifts than at $0<z<2$; ii) the Schechter function slope $\alpha$ is significantly steeper than that displayed by the local stellar mass function, which is both a consequence of the steeper faint end and the absence of a pure exponential decline at the high-mass end; iii) the evolution of the comoving stellar mass density from $z=0$ to 5 can be modelled as $\log_{10} \rho_{M} =-(0.05 \pm 0.09) \, z^2 - (0.22\mp 0.32) \, z + 8.69$. At $3 \leq z<4$, more than 30\% of the $M \gsim 10^{11} \, \rm M_\odot$ galaxies would be missed by optical surveys with $R<27$ or $z<26$. Thus, our study demonstrates the importance of deep mid-IR surveys over large areas to perform a complete census of massive galaxies at high $z$ and trace the early stages of massive galaxy assembly.

\end{abstract}


\section{Introduction}
\label{sec_intro}

The study of massive ($M \gsim 5 \times 10^{10} \, \rm M_\odot$)  galaxies at high ($z>3$) redshifts allows for the investigation of the first epochs of efficient stellar mass assembly, when the Universe was less that a few gigayears (Gyr) old. It is now accepted that around 20-40\% of the massive galaxies we know today were already in place by $z\sim2$  (Fontana et al.~2004; Caputi et al.~2005, 2006a; Daddi et al.~2005; Labb\'e et al.~2005; Saracco et al.~2005; Papovich et al.~2006; Arnouts et al.~2007; Pozzetti et al.~2007; Wuyts et al.~2009; Ilbert et al.~2010). Above redshift $z\sim3$, massive galaxies are more difficult to find (e.g. McLure et al.~2006; Kodama et al.~2007; Rodighiero et al.~2007) and they become very rare by $z\sim5-6$ (e.g. Dunlop, Cirasuolo \& McLure~2007).

The galaxy populations discovered at $z>6$ seem to almost exclusively consist of intermediate or low mass galaxies ($M \lsim  10^{10} \, \rm M_\odot$), and their current study is mainly focused on constraining the epoch and sources of reionisation (e.g. Bunker et al.~2004; Bouwens et al.~2008;  McLure et al.~2009, 2010; Oesch et al.~2010). In fact, with the current instrumentation, only the rest-frame ultra-violet (UV) emission can be observed for the vast majority of these galaxies, which is used to estimate their levels of star formation and ability to produce and liberate sufficient Lyman-$\alpha$ photons to ionise the intergalactic medium. Any constraints on the stellar masses of $z>6$ galaxies are still poor, due to the  lack of sensitive data at wavelengths that map the galaxy rest-frame near-infrared (IR) at these redshifts (although see e.g. Labb\'e et al.~2010 for an attempt). Nevertheless, different pieces of observational evidence suggest that massive galaxies as a significant population only appear at later times.

Cosmological models of galaxy formation predict that massive galaxies can be quickly formed at high-$z$ in the high-density fluctuations of the matter density field (Cole \& Kaiser~1989; Mo \& White~1996). Thus, determining the first epoch of appearance and subsequent rise in the number density of massive galaxies with redshift constitutes a very important constraint on galaxy formation models. Investigating the period elapsed between redshifts $z\sim3$ and 6-7  is fundamental for this purpose, as it connects the earliest stages of galaxy formation after the epoch of reionisation, with the better-studied period of galaxy evolution at $1<z<3$, where a substantial population of galaxies are already massive and host very intense star formation and quasar activity.

A global picture of the evolution of galaxy formation and growth can be obtained through the study of the galaxy stellar mass function at different redshifts. At low redshifts, the shape of the galaxy mass function is known down to low mass limits ($M\sim 10^8 \, \rm M_\odot$; e.g. Cole et al.~2001; Baldry, Glazebrook \& Driver~2008)  and is well-fitted by a double Schechter (1976) function (Baldry et al.~2008; Bolzonella et al.~2010; Pozzetti et al.~2010). Peng et al.~(2010) proposed that the Schechter-function shape observed for the galaxy stellar mass function up to redshift $z\sim2$ can be explained as a consequence of mass-driven star-formation quenching proceeding proportionally to the galaxy star-formation rate (SFR) in $M \gsim M^\ast$ galaxies. This mechanism could be at play since earlier times, as preliminary determinations over small areas of the sky indicate that the Schechter functional form could be suitable to describe the bright end of the galaxy stellar mass function up to $z\sim3.5$ (e.g. Fontana et al.~2006; Kajisawa et al.~2009).

Selecting galaxies by their rest-frame near-IR light constitutes a good proxy for a stellar mass selection. Rest-frame near-IR wavelengths are relatively unaffected by dust, and the corresponding mass-to-light ratios have much smaller variations with galaxy age than at shorter wavelengths. For galaxies at $z\geq3$, the rest-frame near-IR light is shifted into observed mid-IR wavelengths. The Infrared Array Camera (IRAC; Fazio et al.~2004) on board the {\em Spitzer Space Telescope} (Werner et al.~2004)  is currently the most suitable instrument to conduct a mass-selected galaxy survey at high redshifts. IRAC has operated at wavelengths  3.6, 4.5, 5.8 and 8.0 $\rm \mu m$ until the end of the  {\em Spitzer} cryogenic mission, and still operates in the two shortest-wavelength channels in the on-going warm campaign. {\em Spitzer} data constitute the last opportunity to conduct such mid-IR galaxy surveys until the advent of the {\em James Webb Space Telescope (JWST)} after 2014.

The {\em Spitzer} Ultra-Deep Survey (SpUDS; P.I. J. Dunlop) is a Legacy Program that has provided IRAC (Fazio et al.~2004) and  Multiband Photometer for {\em Spitzer} (MIPS; Rieke et al.~2004) imaging over more than 1~deg$^2$ centred on the UKIDSS UDS field (P.I. O. Almaini). The UDS is one of five on-going surveys which comprise the UKIDSS, and is characterised by the existence of deep UV through $K$-band ground-based photometric data over an overlapping area of $\sim 0.60$ deg$^2$. The availability of deep and homogeneous-quality multi-wavelength data in this region makes of the UKIDSS UDS one of the most suitable surveys to perform galaxy evolution studies at high $z$, as demonstrated by several works (e.g. McLure et al.~2006, 2009; Dunne et al.~2009; Cirasuolo et al.~2007, 2010; Hartley et al.~2010; Ono et al.~2010).

In this work, we present the results of an IRAC \ffm selection and multi-wavelength analysis of galaxies at $z\geq 3$ in the UKIDSS UDS field. In Section \S\ref{sec_data}, we describe the different analysed datasets. In Section \S\ref{sec_sample}, we explain in detail our sample selection based on the photometric redshifts of the entire \ffm galaxy catalogue. In Section \S\ref{sec_massfun}, we compute the galaxy stellar mass function  and corresponding stellar mass densities at $3 \leq z<5$. Later, in Section \S\ref{sec_zge5}, we discuss the reliability of our IRAC-selected galaxy candidates at $z>5$. In Section \S\ref{sec_optsel}, we quantify the fraction of  our $3 \leq z <5$ that would be missed by typical deep optical selections. Finally, in Section \S\ref{sec_concl}, we summarise our findings and present our conclusions. We adopt throughout a cosmology with  $\rm H_0=70 \,{\rm km \, s^{-1} Mpc^{-1}}$, $\rm \Omega_M=0.3$ and $\rm \Omega_\Lambda=0.7$.   All quoted magnitudes and colours are total and refer to the AB system (Oke \& Gunn~1983), unless otherwise stated.

\section[]{Datasets}
\label{sec_data}

The UKIDSS (Lawrence et al.~2007) UDS has been conducted over a field centred on  RA=02:17:48 and DEC=-05:05:57 (J2000),  and benefits from a range of multi-wavelength data from X-rays to radio wavelengths. The field is defined by its UKIRT Wide Field Camera (WFCAM) coverage in the $J, H$ and $K$ bands, which is still in progress. The data used in this work corresponds to the fifth data release (DR5), and reach 5$\sigma$ depths of 24.0, 23.7 and 23.9 (2-arcsec-diameter aperture) magnitudes in the $J, H$ and $K$ bands, respectively. 

The UDS field has been observed at optical wavelengths with SuprimeCam on Subaru (Furusawa et al.~2008). This has provided  $B, V, R, i$ and $z$-band data, with corresponding 3$\sigma$ limit (2-arcsec-diameter aperture) magnitudes $B=28.4$, $V=27.8$, $R=27.7$, $i=27.7$ and $z=26.6$. Also, complementing $U$-band data (P.I. O. Almaini) have been obtained with Megacam on the Canada-France Hawaii Telescope (CFHT).

At mid-IR wavelengths, the UDS field has been observed with IRAC and MIPS for a total time of 124 and 168 hours, respectively, as part of a {\em Spitzer} cycle-4 Legacy Program (SpUDS; P.I. J. Dunlop). IRAC observations have been carried out with the four IRAC filters, i.e. at 3.6, 4.5, 5.8 and 8.0 $\rm \mu m$. One of the key science drivers of the SpUDS program is the possibility of exploiting the combined power of optical/near-IR and mid-IR data to study the evolution of the galaxy stellar mass function at high redshifts.

We have used the {\em Spitzer}/IRAC  \ffm data on the UDS field to select a galaxy catalogue at $z \geq 3$ over a net area of 0.60 deg$^2$ of the UDS field with UV through mid-IR coverage. This is the clean area where all datasets overlap, and excludes all map edges and regions around bright stars. In the next section, we explain in detail the steps of our high-$z$ galaxy selection.

\begin{figure}
\includegraphics[width=75mm]{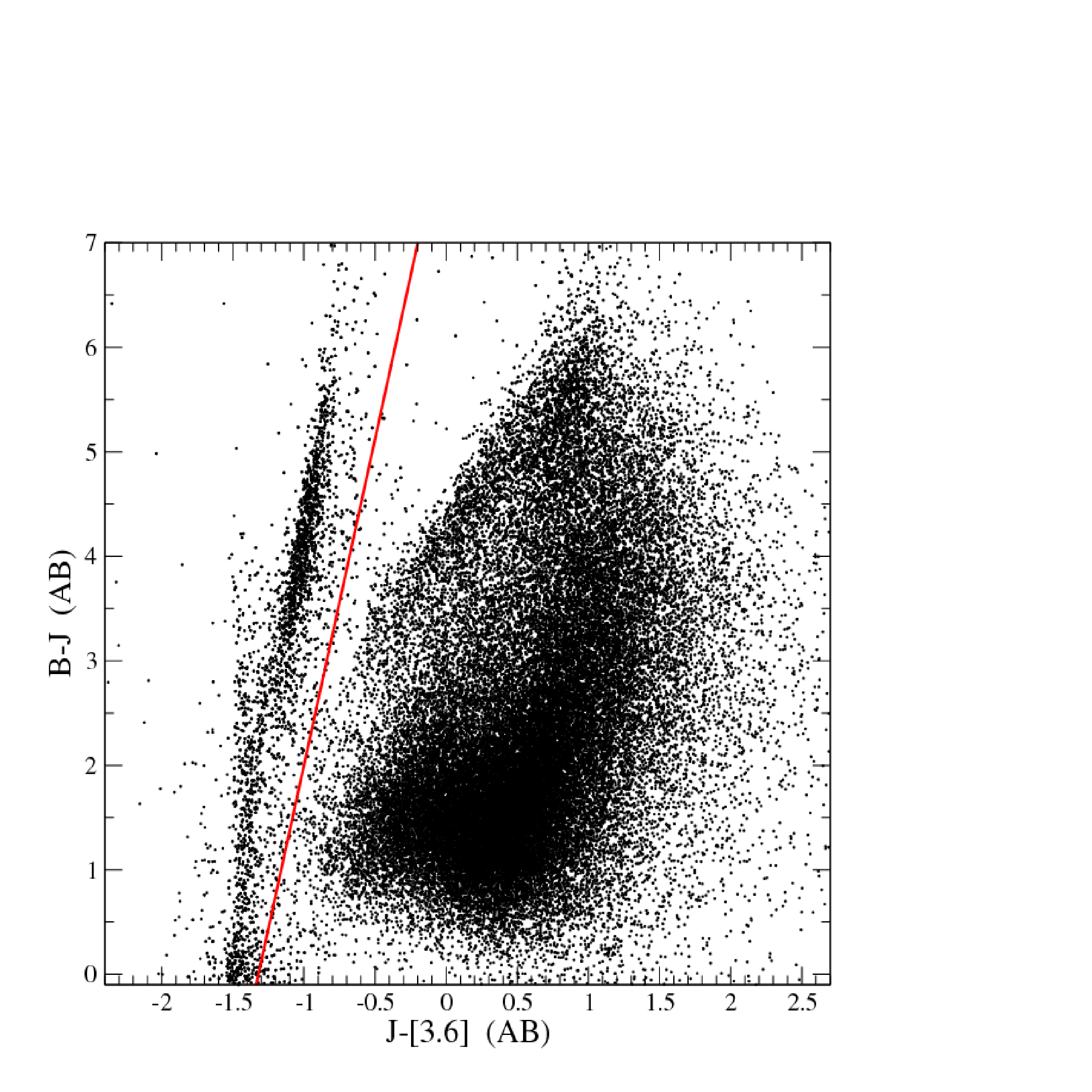}
\caption{The $(B-J)$ versus $(J-[3.6])$ colour-colour diagram for our 52,693 \ffm-selected sources. Stars form a clearly separated sequence on the left-hand side of the diagram.}
\label{fig_stargal}
\end{figure}

\section[]{The S\lowercase{p}UDS IRAC $4.5 \, \rm \mu$\lowercase{m}-selected galaxy sample at \lowercase{$z$}$\geq3$}
\label{sec_sample}

\subsection{Sample selection and multi-wavelength photometry}
\label{sec_sample}

 We extracted a source catalogue from the SpUDS \ffm image using the software SEXTRACTOR (Bertin \& Arnouts~1996) with a `mexhat' kernel. This type of kernel is very efficient in crowded fields, as it facilitates source deblending. Considering only the region overlapping the 3.6 $\rm \mu m$ map, and excluding edges and regions around bright stars, our 4.5 $\rm \mu m$ catalogue contains 67,937 sources.  
 
 We measured aperture photometry for all our sources and obtained aperture corrections using the curve of flux growth for isolated stars in the field. Our derived total \ffm magnitudes --referenced as [4.5] hereafter-- correspond to measured 4-arcsec-diameter aperture magnitudes corrected by a constant -0.31 mag. This aperture size is usual for IRAC photometry (see e.g.  Ilbert et al.~2010), as it constitutes a good balance between directly measuring most of the source encircled energy and minimising contamination from close neighbours (the IRAC point-spread function full width half maximum is $\sim$1.9 arcsec at \ffm).

 We performed simulations to assess the completeness and reliability of our \ffm catalogue. To test completeness, we used the IRAF task `gallist' to generate a list of 50,000 artificial objects following a power-law distribution between magnitudes 18 and 26. We then created a set of 100 mock maps based on the real \ffm image, in each of which we have randomly inserted 500 of the artificial objects (using `mkobjects' in IRAF).  We then ran SExtractor on each of these mock maps with the same configuration file used for the real image, and checked the fraction of artificial sources recovered as a function of magnitude. Through this procedure, we determined that our \ffm catalogue is 80\% and 50\%  complete to magnitudes [4.5]=22.4 and 24.0, respectively. 
 
 We tested the reliability of our catalogue by repeating the source extraction procedure on the negative of the \ffm image, and considering the fraction of negative sources versus magnitude. At [4.5]=22.4 mag, the percentage of spurious sources is below 0.5\%. At fainter magnitudes [4.5]=23.5-24.0 mag, this percentage rises to around 10\%. However, after imposing that the \ffm sources have a counterpart in the independently extracted $K$-band catalogue (see below), the fraction of spurious sources becomes negligible even at such faint magnitudes.

 We measured 3.6 $\rm \mu m$ aperture photometry for all the \ffm-selected sources running Sextractor in dual-image mode. The derived total 3.6 $\rm \mu m$ magnitudes correspond to the measured 4-arcsec-diameter aperture magnitudes corrected by a constant -0.27 mag (as also determined through the curve of flux growth of isolated stars). 
 
 To compile the corresponding UV through near-IR photometry for our galaxies, we extracted an independent catalogue based on the UDS $K$-band image, and ran SExtractor on dual-image mode on the $U, B, V, R, i, z, J$ and $H$-band maps,  using the position of the $K$-band sources. In these bands, we obtained total magnitudes from aperture-corrected 2-arcsec aperture magnitudes in all cases. All magnitudes have been corrected for galactic extinction.

 We finally cross-correlated the \ffm catalogue (that included 3.6 $\rm \mu m$ photometry) with the $K$-band catalogue (that contained $U$-band through $K$-band photometry), with a matching radius $r=1.5$ arcsec. The final overlapping area of all our datasets is 0.60 deg$^2$. Our \ffm catalogue with  $K$-band counterparts over this area contains 52,693 sources. 
 
 We note that the depth of the near-IR images  matches very well the depth of the IRAC data in the UDS. Within the clean overlapping area of 0.60 deg$^2$, the $K$-band catalogue allows us to identify more than 98\% of the \ffm sources with [4.5]$<22.4$ mag. For the deeper [4.5]$<24.0$ mag catalogue, the percentage of identifications is 92\%. Our reliability tests performed on the  \ffm catalogues suggest that most of the remaining unidentified sources are likely to be spurious IRAC sources.
 
 We excluded galactic stars from our sample via a colour-colour diagram. As discussed by McLure et al.~(2009), the use of the SExtractor stellarity parameter CLASS\_STAR alone is not a secure way to segregate stars from high-$z$ galaxies when using ground-based data, as some of the galaxies are compact and could also have large stellarity parameters ($\rm CLASS\_STAR>0.8-0.9$). Instead, colour segregation is much more reliable. Fig. \ref{fig_stargal}  shows that stars form a separate sequence in the $(B-J)$ versus $(J-[3.6])$ colour-colour diagram. Through this colour diagnostic, we determined that 2372 out of our 52,693 sources are galactic stars. Note, however, that this colour-colour diagram cannot segregate red dwarf stars, which are a potential source of contamination for high-$z$ galaxy samples (cf. Section \S\ref{sec_zge5}).
 
Basically all of the 2372 colour-segregated objects have $\rm CLASS\_STAR>0.8$, but they constitute less than a half of the total number of sources with $\rm CLASS\_STAR>0.8$ within our sample (in our case, we measured the $\rm CLASS\_STAR$ parameter on the $K$-band images). After the star separation, the final catalogue that we considered for further analysis contains 50,321 galaxies.

\begin{figure}
\begin{center}
\includegraphics[width=70mm]{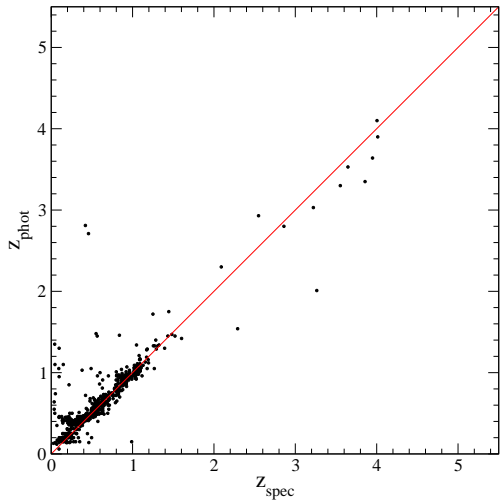}
\caption{Photometric versus spectroscopic redshifts for 733 galaxies in our \ffm-selected sample. The overall quality of our photometric estimates is $\rm \sigma[|z_{\rm phot}-z_{\rm spec}|/(1+z_{\rm spec})]=0.05$, after excluding catastrophic outliers (3\% of the data points).}
\label{fig_zphzsp}
\end{center}
\end{figure}

\subsection{Photometric redshifts and selection of the $z\geq3$ sample}
\label{sec_zphot}

We derived photometric redshifts $z_{\rm phot}$ for our 50,321 \ffm-selected galaxies making use of the photometric data in 11 broad bands ($U$ through \ffm), and a code largely based on  HYPERZ (Bolzonella, Miralles \& Pell\'o~ 2000; see Cirasuolo et al.~2010 for details). The spectral energy distribution (SED) fitting  has been done using the 2007 version of the Bruzual \& Charlot templates (Bruzual \& Charlot 2003; Bruzual~2007), and applying the Calzetti et al. (2000) reddening law with $0.0\leq A_V \leq 3.0$ to account for internal extinction.

As a diagnostic of the quality of our derived $z_{\rm phot}$, we compared these values with the corresponding real redshifts for a subset of 733 galaxies for which spectroscopic data are available (Fig. \ref{fig_zphzsp}). The overall quality of our photometric redshifts is very good: the mean of the $(z_{\rm phot}-z_{\rm spec})/(1+z_{\rm spec})$ distribution is -0.01 and the dispersion is $\sigma=0.05$ after removing only 3\% of outliers. The final photometric redshifts considered for this diagram incorporate all the further tests performed on the photometric redshifts that are explained below.

Note that, although unfortunately we do not have spectroscopic confirmation for most of the $z\geq 3$ galaxies that we study in this work, the $z_{\rm phot}$ versus $z_{\rm spec}$ diagram gives us confidence that our discrimination of lower-redshift galaxies is reliable.

We used the  $z_{\rm phot} - z_{\rm spec}$ diagnostics to iteratively correct the zero-points of our photometric data in different bands.   The final systematic zero-point corrections we applied to our photometry  is -0.25 mag in the $U$ band, and -0.10 mag for the $J,H$ and $K$ bands.

Our photometric redshifts are the first criterion to select the $z\geq3$ galaxy sample. Based solely on the primary solutions (i.e. the photometric redshift corresponding to the minimum $\chi^2$ value for each galaxy), our sample contains 1608 $z\geq3$ candidates. However,  no galaxy should have any significant flux  blueward of the Lyman-$\alpha$ limit at rest $\rm \lambda_{rest}=912 \, \AA$. So, as a further test, we checked this condition on our $z\geq3$ candidates. 

At $z>3.6$, the Lyman-$\alpha$ limit is redward of the $U$-band filter. Thus, for
accepting a galaxy candidate at $z>3.6$, we imposed that it should be a $U$-band dropout (i.e. it has a $U$-band magnitude below the $2 \sigma$ detection level of our catalogue). Following the same criterion, for accepting a galaxy at $z>4.5$, we imposed that it should be undetected in both the $U$ and $B$ bands. After these restrictions, a total of 1544 galaxies remain in our $z\geq3$ sample.

On the other hand, the SED-fitting procedure yields the $\chi^2$ distribution as a function of redshift for each galaxy, and with this we can assess the probability $P(z)$ that each of the remaining candidates is truly at $z\geq3$. For 244 of them, the SED fitting indicates that secondary, lower redshift ($z_{\rm sec}<3$) solutions cannot be discarded within $3\sigma$ confidence (i.e. $\rm \chi^2_{sec.} - \chi^2_{prim}< 9$).  Comparison of photometric with spectroscopic redshifts for a subset of these objects indicates that the secondary, lower-redshift solution is the correct one in these cases  (in the plot shown in Fig. \ref{fig_zphzsp}, the percentage of catastrophic outliers would more than double if we considered the primary photometric redshifts for these objects). Therefore, we decided to exclude from our final sample these 244 galaxies for which secondary $z<3$ photometric redshift solutions cannot be discarded within $3\sigma$ confidence. This is a conservative criterion that is necessary to guarantee that we do our science analysis on a secure high-$z$ sample.

\begin{figure}
\begin{center}
\includegraphics[width=70mm]{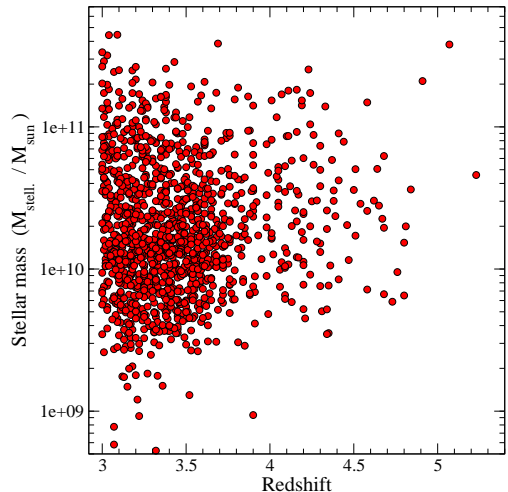}
\caption{Estimated stellar masses versus redshifts for the 1292 galaxies in our  $z\geq3$ sample. A Salpeter~(1955) IMF over stellar masses $M=(0.1-100) \, \rm M_\odot$ is assumed.}
\label{fig_stmvsz}
\end{center}
\end{figure}

The 244 rejected sources constitute $\sim 15\%$ of our initial $z\geq3$ candidate sample. We analyse the potential impact that these sources would have in the galaxy stellar mass function at $3\leq z <5$ in Appendix A.

Our remaining  $z\geq3$ sample contains 1300 galaxies, including  ten sources that are $z>5$  candidates. We separately discuss these ten sources in Section \S\ref{sec_zge5}, but we anticipate that 8 out of these 10 candidates are likely not genuine $z>5$ galaxies. Our final, secure $z \geq 3$ sample contains 1292 galaxies. A total of 1215 out of these 1292 galaxies are within the 50\% completeness limit of our \ffm catalogue, i.e. have $\rm [4.5] \leq 24.0$ mag.

\section{The galaxy stellar mass function at $3\leq\lowercase{z}<5$}
\label{sec_massfun}

\subsection{The $\rm 1/V_{max}$ method and the maximum likelihood analysis}
\label{sec_vmaxml}

We derived stellar mass estimates for each of our galaxies from the best SED fitting performed with the 2007 version of the Bruzual \& Charlot template library.  We used a set of templates corresponding to a single stellar population, and a collection of exponentially declining star formation histories with characteristic times ranging between $\tau=0.01$ and 5 Gyr, all of them with a solar metallicity. The considered age grid was such that the maximum possible age would not exceed the age of the Universe at the redshift of each galaxy. We convolved each template with the Calzetti et al. (2000) reddening law with $0.0\leq A_V \leq 3.0$ to account for internal extinction. 

The UDS multi-wavelength photometry and, particularly, the availability of IRAC data that sample the rest-frame near-IR light of galaxies up to $z>4$, allow us to derive reliable stellar mass estimates for all our sources. In all cases, the stellar masses are based on the best-fitting SED normalised between the corresponding $K$ and [4.5] band magnitudes of each galaxy. This produces less uncertainties in the stellar mass estimates than allowing the SED to be adjusted with the galaxy UV/optical light (cf. Section \ref{sec_edd}).  The derived stellar masses of our sources versus redshift are shown in Fig. \ref{fig_stmvsz}. We assume a Salpeter~(1955) initial mass function (IMF) over the stellar mass range $M=(0.1-100) \, \rm M_\odot$.

\begin{figure*}
\begin{center}
\includegraphics[width=180mm]{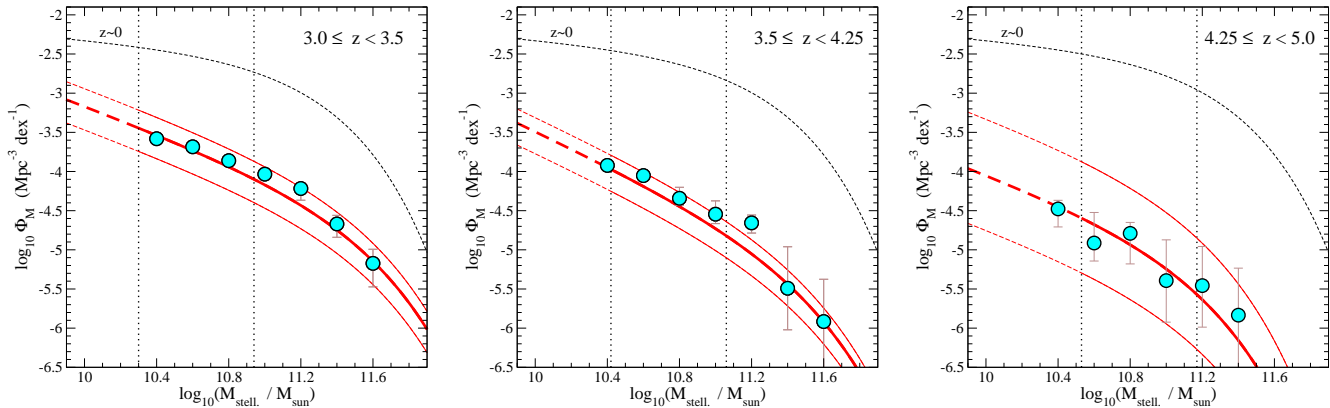}
\caption{The galaxy stellar mass function in three redshift bins between $z=3$ and 5. Circles show the results of the $\rm 1/V_{max}$ method.  The error bars include Poisson statistics and the effects of the galaxy probability density distributions in redshift space (see text). 
 Vertical dotted lines indicate the strict mass completeness --corresponding to a single stellar population formed at $z\rightarrow \infty$-- that are imposed by the 80\% and 50\%
 completess limits of our catalogue at  [4.5]=22.4 and 24.0 mag, respectively.
The thick curve in each panel represents the mass function calculated with an STY maximum likelihood analysis, assuming a Schechter function (solid and dashed line, above and below the mass completeness limit corresponding to [4.5]=24.0 mag, respectively). The accompanying thin lines indicate  the envelope of all possible curves within the 1$\sigma$ confidence limits (see Fig. \ref{fig_mlconf}). The dashed line  with the label $z\sim0$ corresponds to the local stellar mass function computed by Cole et al.~(2001).}
\label{fig_mf3z5} 
\end{center}
\end{figure*}

\begin{table*}
\centering
\caption{The galaxy stellar mass function computed with the $\rm 1/V_{max}$ method.}
\begin{tabular}{c@{\hspace{20mm}}c@{\hspace{15mm}}c@{\hspace{15mm}}c}
\hline
 ${\rm log}_{10} (M / \rm M_\odot)$ &  \multicolumn{3}{c}{${\rm log}_{10} (\rm \Phi_M \, / \, Mpc^{-3} dex^{-1})$}\\
\hline    
   &  $3.0\leq z < 3.5$ & $3.5\leq z < 4.25$ & $4.25\leq z < 5.0$\\
\hline
 10.40  &  $-3.58^{+0.04}_{-0.05}$ &  $-3.92^{+0.05}_{-0.06}$ & $-4.48^{+0.11}_{-0.23}$ \\
 10.60  &  $-3.69^{+0.04}_{-0.05}$ &  $-4.05^{+0.06}_{-0.07}$ & $-4.91^{+0.39}_{-0.23}$\\
 10.80  &  $-3.86^{+0.05}_{-0.06}$ &  $-4.34^{+0.14}_{-0.09}$ & $-4.79^{+0.14}_{-0.39}$\\
 11.00  &  $-4.03^{+0.06}_{-0.07}$ &  $-4.55^{+0.17}_{-0.12}$ & $-5.39^{+0.52}_{-0.53}$\\
 11.20  &  $-4.22^{+0.07}_{-0.15}$ &  $-4.66^{+0.10}_{-0.13}$ & $-5.46^{+0.50}_{-0.53}$\\
 11.40  &  $-4.67^{+0.11}_{-0.17}$ &  $-5.49^{+0.53}_{-0.53}$ & $-5.83^{+0.60}_{-0.76}$\\
 11.60  &  $-5.17^{+0.18}_{-0.30}$ &  $-5.92^{+0.54}_{-0.76}$ & ...\\
\hline		
\end{tabular}
\label{table_vmax}
\end{table*}

For the purpose of the stellar mass function calculation, we considered only the 1213 galaxies of our sample with $\rm [4.5] \leq 24.0$ mag at $3\leq z<5$, and divided the sample into the following redshift bins: $3.0\leq z<3.5$, $3.5\leq z<4.25$ and $4.25\leq z<5.0$, which sample comparable comoving volumes.

We first computed the galaxy stellar mass function using the $\rm 1/V_{max}$ method (Schmidt~1968).   Although this technique involves data binning, it has the advantage of  being free of any parameter dependence or model assumptions. Besides, the normalization is directly obtained from the galaxy counts in each stellar mass bin.  

The comoving volume $\rm V_{max}$ considered for each galaxy depends on the maximum redshift $z_{\rm max}$ at which the galaxy can be observed, given the flux limit of the survey ($\rm [4.5] \leq 24.0$ in our case). If this redshift is larger than the maximum redshift of the corresponding redshift bin $z_{\rm max} \geq z_{\rm max(bin)}$, then $\rm V_{max}=V_{max(bin)}-V_{min(bin)}$. Instead, if $z_{\rm max} < z_{\rm max(bin)}$, then $\rm V_{max}=V_{{\mathit z}_{max}}-V_{min(bin)}$. To account for the sample incompleteness, we weighed each galaxy by a correction factor, depending on its \ffm magnitude.

Our results for the stellar mass function calculated with the $\rm 1/V_{max}$ method are shown in Fig. \ref{fig_mf3z5} (circles) and its values given in Table \ref{table_vmax}. The 80 and 50\% strict stellar-mass completeness limits, corresponding to a single stellar population formed at $z\rightarrow \infty$ with the catalogue limiting magnitudes $\rm [4.5]=22.4$ and 24.0 mag, respectively, are indicated with vertical dotted lines in the three panels of Fig. \ref{fig_mf3z5}.

The error bars of our stellar mass function are based on the result of Monte Carlo simulations and incorporate the full probability density distribution $P(z)$  of each galaxy in redshift space at $z\geq3$.

\begin{figure*}
\begin{center}
\includegraphics[width=150mm]{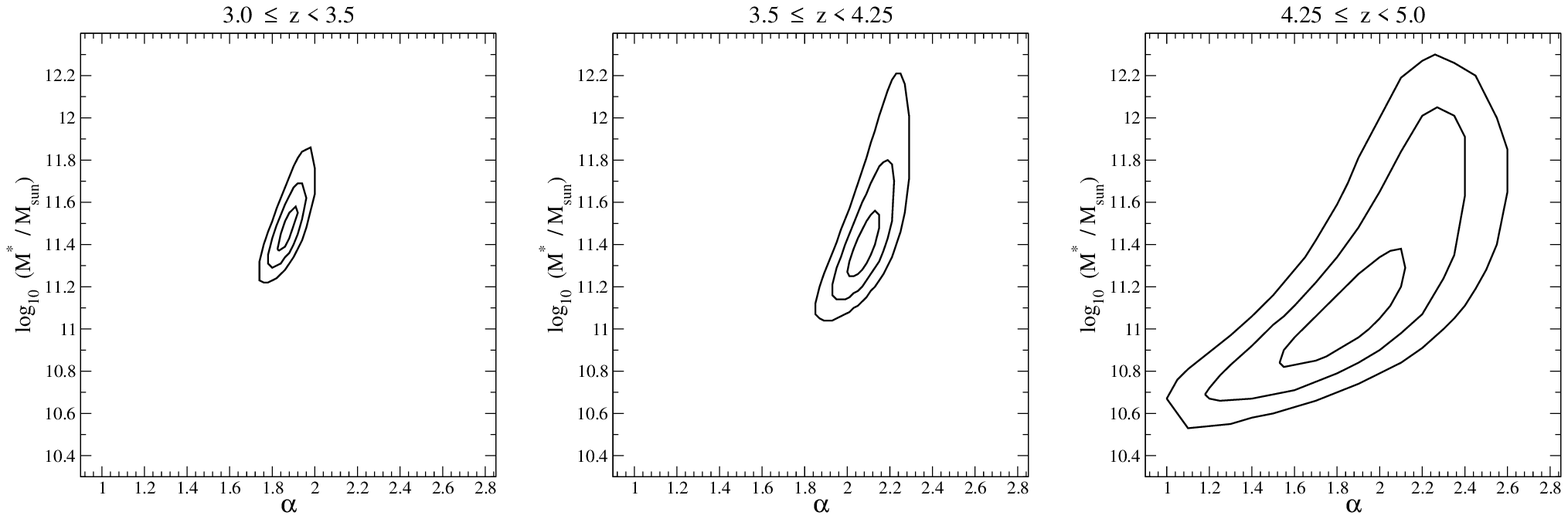}
\caption{The 1, 2 and 3$\sigma$ confidence levels for the maximum-likelihood free parameters.}
\label{fig_mlconf} 
\end{center}
\end{figure*}

\begin{table*}
\centering
\caption{Maximum likelihood parameter values for the stellar mass function computed with the STY analysis assuming a Schechter function. The last column lists the values of the comoving stellar mass density obtained by integrating $M \,\Phi(M) \, d\log_{10}(M)$ over  stellar masses $\log_{10}(M)=8.0$ to 13.}
\begin{tabular}{l@{\hspace{10mm}}c@{\hspace{10mm}}c@{\hspace{10mm}}c@{\hspace{10mm}}c}
\hline
 Redshift &  $\alpha$ &  $M^\ast \, \rm (M_\odot)$ &  $\Phi^\ast \, \rm (Mpc^{-3} \, dex^{-1})$ & $\rho_M (\rm M_\odot \, Mpc^{-3})$\\
\hline 
$3.0\leq z < 3.5$ & $1.86^{+0.05}_{-0.04}$ & $(2.82^{+1.02}_{-0.48})\times 10^{11}$ & $(3.95^{+1.65}_{-1.42})\times 10^{-5}$ & $(2.10^{+0.19}_{-0.08})\times 10^7$\\
$3.5\leq z < 4.25$ & $2.07^{+0.08}_{-0.07}$ & $(2.34^{+1.29}_{-0.56})\times 10^{11}$ & 
$(1.15^{+0.61}_{-0.55})\times 10^{-5}$ & $(1.12^{+0.32}_{-0.23})\times 10^7$ \\
$4.25\leq z < 5.0$ & $1.85^{+0.27}_{-0.32}$ & $(1.15^{+1.25}_{-0.49})\times 10^{11}$ &
$(1.21^{+5.21}_{-0.97})\times 10^{-5}$ & $(2.36^{+0.89}_{-1.68})\times 10^6$\\
\hline		
\end{tabular}
\label{table_ml}
\end{table*}

We used the $P(z)$ function of each of the 1213 galaxies involved in the computation of the stellar mass function to create 1000 mock catalogues. Each of these catalogues also contains 1213 galaxies with the redshifts randomly assigned with a probability given by the $P(z)$ function of each source, and constrained by the dropout criteria discussed in Section \ref{sec_zphot}.   A new stellar mass is derived so as to be consistent with the new random redshift and best-fitting SED in each case. We then re-computed  the stellar mass function with the $\rm 1/V_{max}$ method for each mock catalogue, and derived 1$\sigma$ errors for the values originally obtained on the real sample. Poisson errors are quoted instead when they are larger than the errors derived from the Monte Carlo simulations.
 
 The comparison of the values given in Table \ref{table_vmax} shows a significant decrease of the galaxy stellar mass function with increasing redshift within the $3 \leq z < 5$ interval. For all stellar masses, we find that the number of galaxies decreases by a factor of $\sim 10$ between  $3.0\leq z < 3.5$ and $4.25\leq z < 5.0$.  This is particularly the case for the three largest stellar mass bins, in which our sample is 80\% complete at all these redshifts. Thus, we can conclude that the decrease in the mass-function is a real effect.

We also performed a second, independent computation of the galaxy stellar mass function applying the STY (Sandage, Tammann \& Yahil~1979) maximum likelihood  analysis.  This is a parametric technique that assumes that the shape of the mass function is known. However, in contrast to the $\rm 1/V_{max}$ method, this method does not involve data binning and does not contain any implicit assumption on a uniform spatial distribution of galaxies. The maximum likelihood analysis provides a direct calculation of the stellar mass function, i.e. it does not constitute a fitting procedure of the stellar mass function obtained with the $\rm 1/V_{max}$ method.

The corresponding maximum likelihood estimator reads:

\begin{eqnarray}
\lefteqn{\mathcal L [s_k | (z_i, M_i)_{i=1,...,N}] =} \nonumber\\
&& = \prod_{i=1}^N \, \left[\frac{\Phi(s_k, M)}{\int_{\log_{10} (M_{0}^{\mathnormal i})}^{+ \infty}\Phi(s_k, M) \, d\log_{10}(M)}\right]^{w_i} \!\!\!\!\!\!,
\end{eqnarray} 

\noindent where the product is made over the $i=1,...,N$ galaxies of the sample.  $\Phi(s_k, M)$ is the adopted functional form for the stellar mass function, which depends on the stellar mass $M$ and the parameters $s_k$. $M_{0}^{\mathnormal i}$ is the minimum stellar mass at which  the $i$-th galaxy would be observable, given its redshift $z_i$ and the flux limit of the survey. The weighting factors $w_i$ allow to take into account sample completeness corrections. By maximizing $\mathcal L$ (or, for simplicity, its logarithm), one can obtain the values of the parameters $s_k$ yielding the maximum likelihood. The normalization factor $\Phi^\ast$ is recovered after the maximization, by integrating the obtained  maximum-likelihood mass function without normalization in the completeness range of stellar masses of the survey, and making it equal to the number density of observed galaxies.

For the shape of the stellar mass function, we assumed the functional form proposed by Schechter~(1976):
 
\begin{equation}
\label{eq-sch}
\Phi(M) \, d\log_{10}(M)=\Phi^\ast \, \left(\frac{M}{M^\ast}\right)^{1-\alpha} \!\!\!\!\!\times \, \exp\left(-\frac{M}{M^\ast}\right) \, d\log_{10}(M),
\end{equation}

\noindent where $\alpha$ and $M^\ast$ are free parameters. The Schechter functional form implies that the value of $\alpha$ dominates the shape of the resulting stellar mass function at $M \lsim M^\ast$. As we mentioned in Section  \S\ref{sec_intro}, this functional form well describes the shape of the stellar mass function up to at least $z\sim3$. We briefly discuss the results of exploring other functional forms  later in this section.

The thick curves in Fig. \ref{fig_mf3z5} show our maximum-likelihood stellar mass function obtained with the STY method in each redshift bin. The best $\alpha$ and $M^\ast$ parameter values (i.e. those yielding the maximum likelihood in each case) are given in Table \ref{table_ml}, and the 1, 2 and 3$\sigma$ confidence levels on these parameters are plotted in Fig. \ref{fig_mlconf}. These confidence levels assume that the likelihood function follows a Gaussian distribution around the maximum, i.e. they are determined by $\Delta \ln (\mathcal L)= -0.5, -2.0$ and -4.5, respectively.

The random errors on $\alpha$, $M^\ast$, and $\Phi^\ast$ produced by the uncertainties in the photometric redshifts --as computed with the mock catalogues created following the $P(z)$ distribution for each galaxy-- are contained within the confidence levels shown in Fig. \ref{fig_mlconf}. The $z_{phot}$  uncertainties can also produce non-negligible systematic effects on the galaxy stellar mass function, particularly the so-called  Eddington bias (Eddington~1913), which we discuss in detail in Section \ref{sec_edd}.

From the results of our maximum likelihood analysis, we find the following. 
On the one hand, the best values of $\alpha$ in the three analysed redshift bins are considerably higher  than the local mass-function slope $\alpha=1.18\pm0.03$ (Cole et al.~2001).  We observe this steepening of the stellar mass function even when the flux limit of our \ffm catalogue mainly restricts our analysis to the region around and above $M^\ast$. In fact, this is not surprising: although it is true that the value of $\alpha$ determines the slope of the stellar mass function in the faint end, it is also important for the bright end, as it modulates the exponential decline of the Schechter function at $M \gsim M^\ast$(cf. eq. \ref{eq-sch}). Put another way, a flat value $\alpha \sim 1$ will only be possible if the stellar mass function decreases exponentially at  $M \gsim M^\ast$.

Our steep $\alpha$ values are closer to those that best describe the optical luminosity function at high-$z$ (e.g. McLure et al.~2009). Also, Fontana et al.~(2006) and Kajisawa et al.~(2009) found that the slope of the stellar mass function increased with redshift up to $z\sim 4$. 

At $3<z<4$, Fontana et al.~(2006)  found $\alpha \approx 1.5$ and $M^\ast \approx 8\times 10^{10} \, \rm M_\odot$. Note that, although these values are lower than those we find  in this work, the agreement between our stellar mass function and that of Fontana et al.~(2006) is very good (cf. Section \S\ref{sec_compoth}). This is because the values of the parameters $\alpha$ and $M^\ast$ are coupled, and different combinations of the two can produce a similar shape for the stellar mass function around and above $M \sim M^\ast$ (where current observations actually constrain the stellar mass function). 

At redshifts $2.5<z<3.5$, Kajisawa et al.~(2009) obtained $\alpha \approx 1.75$, a similar value to that we find here. Their corresponding characteristic stellar mass is $M^\ast \approx 2.6\times 10^{11} \, \rm M_\odot$.

On the other hand, we find that our maximum likelihood values for $M^\ast$ and $\Phi^\ast$ decrease with redshift within the range $3 \leq z<5$, in consistency with our  $\rm 1/V_{max}$-method results. The comparison of $M^\ast$ and $\Phi^\ast$ is straightforward in the $3.0 \leq z < 3.5$ and $4.25 \leq z < 5.0$ redshift bins, as the $\alpha$ values are very similar. Between these two redshift bins,   $M^\ast$ decreases to less than a half, and $\Phi^\ast$ becomes around three times smaller. Therefore, our results indicate a fast evolution of the stellar mass function in the short ($\sim$1 Gyr) elapsed time between redshifts $z=3$ and 5.

As an alternative to the Schechter function, we tried a double-exponential form in the maximum likelihood analysis.  The two functional forms can only be distinguished in practice when the stellar mass function is well constrained at the very bright end ($M \gg M^\ast$; cf. Caputi et al.~2007 for a discussion in the case of luminosity functions).  In our present analysis, we find that the maximum likelihoods obtained with either the Schechter or the double-exponential functions are consistent within 3$\sigma$  (the resulting stellar-mass-function curves are almost identical within the stellar mass range sampled in our catalogue). 

A single power-law can be discarded as the shape of the stellar mass function at $3.0 \leq z < 3.5$ and $3.5 \leq z < 4.25$ with $>6\sigma$ and $>3 \sigma$ confidence, respectively. Instead, we cannot exclude the single power-law shape at $4.25 \leq z < 5.0$, for which the maximum likelihood is within $\sim 2 \sigma$ confidence of the Schechter function case. It is likely that this effect is produced simply because we might not be well sampling the $M<M^\ast$ regime of the stellar mass function at these redshifts. However, this could alternatively be suggesting that the shape of the stellar mass function is actually changing around this cosmic epoch.  Deeper mid-IR observations that allow a better sampling of the faint end of the stellar mass function are necessary to confirm that this is a real effect.

\subsection{Tests on the faint-end slope $\alpha$}
\label{sec_alphatest}

\begin{figure}
\begin{center}
\includegraphics[width=60mm]{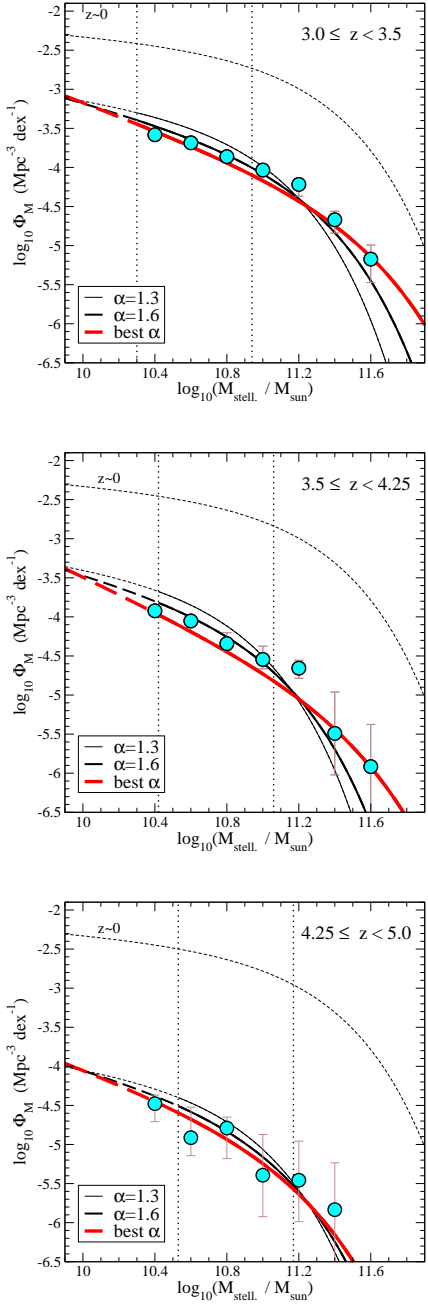}
\caption{Comparison of the galaxy stellar mass function obtained with the ML analysis at different redshifts,  assuming a Schechter function with fixed slope $\alpha=1.3$, 1.6 and a free $\alpha$ value.  Circles show the results of the $\rm 1/V_{max}$ method (cf. Fig. \ref{fig_mf3z5}).}
\label{fig_alph16} 
\end{center}
\end{figure}

We performed a few tests on our galaxy stellar mass function to assess the robustness of the large values we get for the faint-end slope parameter $\alpha$.

Firstly, we simply repeated our ML analysis fixing the value of $\alpha$ to different smaller values.  Fig.~\ref{fig_alph16} compares the galaxy stellar mass function resulting from alternatively considering fixed $\alpha=1.3$, 1.6, and leaving $\alpha$ as a free parameter (cf. Section \S\ref{sec_vmaxml}).  As before, the normalisation $\Phi^\ast$ in each case has been obtained a posteriori, by imposing that the integral of the stellar mass function is equal to the source number density down to the mass completeness limit. 

Our results show that, at $3.0 \leq z < 3.5$ and  $3.5 \leq z < 4.25$, the best galaxy stellar mass function
that we obtained with a Schechter function with fixed $\alpha=1.6$ is still in a reasonable agreement with the stellar mass function computed with the $\rm 1/V_{max}$ method.  Instead, a fixed $\alpha=1.3$ value 
significantly under-predicts the stellar mass function high-mass end, and at the same time over-predicts the number density of galaxies at intermediate stellar masses $M \sim 2-6 \times 10^{10} \, \rm M_\odot$.
These results indicate that a slope as small as  $\alpha=1.3$ is not adequate to describe the galaxy stellar mass function derived from our datasets.

At $4.25 \leq z < 5.0$, the same effect is present, although the constraints on the stellar mass function provided by our data in this redshift bin are less tight, and the case of a small  $\alpha$ value cannot be rejected. (See that the value $\alpha=1.3$ is contained within the 2$\sigma$  confidence level shown in the right-hand panel of Fig.\ref{fig_mlconf}).

Another test we performed to assess the robustness of a large $\alpha$ value consisted in restricting our ML analysis to our galaxy sample with 80\% completeness, i.e. only to sources with $\rm [4.5] \leq 22.4$ mag. We only performed this test in the $3.0 \leq z <3.5$ redshift bin, where we have a sufficient number of  $\rm [4.5] \leq 22.4$ galaxies as to allow both $\alpha$ and $M^\ast$ to be treated as free parameters.

\begin{figure}
\begin{center}
\includegraphics[width=60mm]{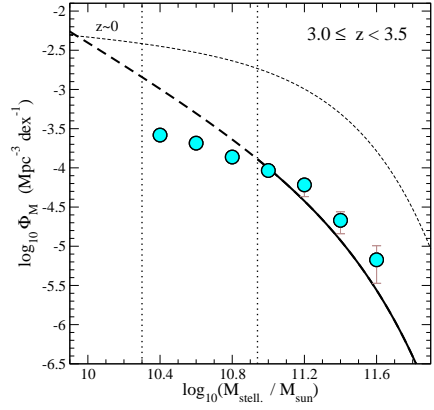}
\caption{The galaxy stellar mass function at $3.0 \leq z <3.5$ obtained by applying the maximum likelihood analysis only to those galaxies with $\rm [4.5] \leq 22.4$ mag. Circles show the results of the $\rm 1/V_{max}$ method for all galaxies down to $\rm [4.5]=24.0$ mag.}
\label{fig_mf224} 
\end{center}
\end{figure}

Figure \ref{fig_mf224} shows our results. Our maximum likelihood analysis applied only to the $\rm [4.5] \leq 22.4$ mag galaxies yields a significantly larger value of the faint-end slope, i.e. $\alpha \approx 2.4$, and $M^\ast \approx 1.8 \times 10^{11} \, \rm M_\odot$. However, we see that the extrapolation of this stellar mass function curve would also overpredict the $\rm 1/V_{max}$ points obtained for stellar masses lower than the strict completeness limit imposed by the $\rm [4.5]=22.4$ magnitude cut.

In conclusion, our tests confirm that the galaxy stellar mass function at $3 \leq z <5$  is characterised by a high value of the slope $\alpha$. The best slope values we obtain from our datasets are $1.8 < \alpha < 2.1$, but in practice somewhat more modest values such as $\alpha=1.6$ could still be adequate. Instead, much smaller $\alpha$ values as those characterising the galaxy stellar mass function at low-$z$ can be clearly discarded.

We note that, in the Schechter function, the $\alpha$ value determines the slope of the stellar mass function at the faint end, but has also the role of modulating the exponential decline at $M \gsim M^\ast$.  Thus, a large $\alpha$ value in the Schechter function is indicating the absence of a pure exponential decline at the high-mass end. This effect is particularly at play within our sample, as it mostly constrains the stellar mass function around and above $M^\ast$ rather than $M \ll M^\ast$.

\subsection{Analysis of the Eddington bias}
\label{sec_edd}

\begin{figure}
\begin{center}
\includegraphics[width=70mm]{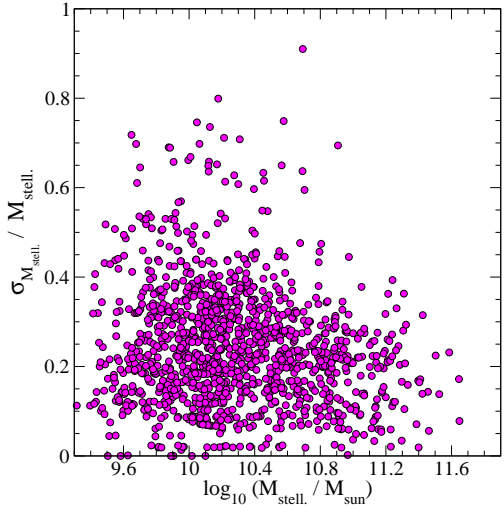}
\caption{The relative stellar mass error versus the estimated stellar mass of each
of our galaxies. These errors consider the parameter degeneracies of the SED fitting at the best-fitting redshift. In all cases, to derive the stellar masses, the SED model is normalised between the $K$ and [4.5] band magnitudes of each source.}
\label{fig_relerrstm} 
\end{center}
\end{figure}
 
\begin{figure}
\begin{center}
\includegraphics[width=60mm]{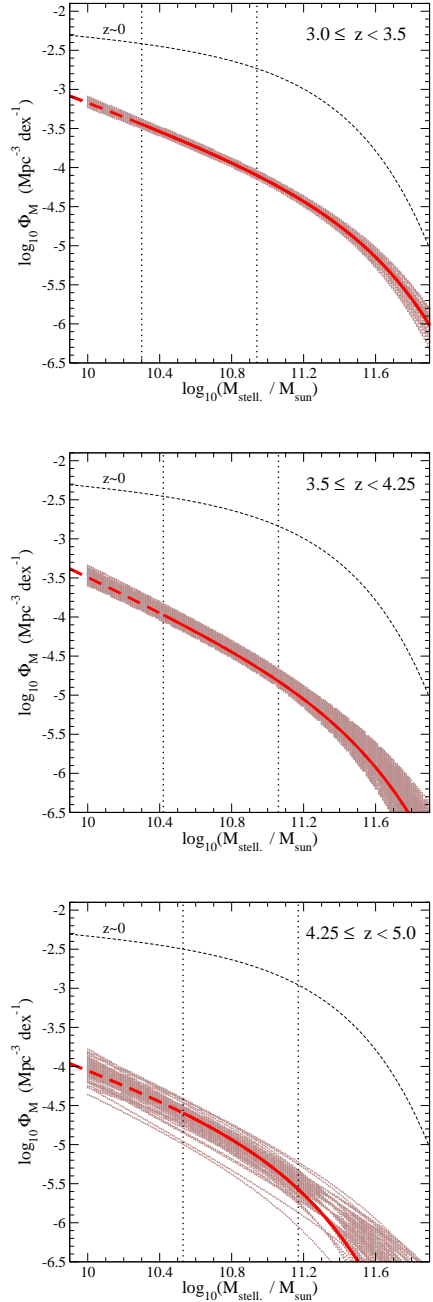}
\caption{The effect of considering the $z_{phot}$ and stellar mass errors of each galaxy on the determination of the galaxy stellar mass function. In each panel, the thick line corresponds to the stellar mass function obtained applying the maximum likelihood analysis on the real sample of 1213 $z\geq3$ galaxies with the original redshifts (cf. Fig. \ref{fig_mf3z5}). The multiple thin lines correspond to 100 Monte Carlo realizations, with the  maximum likelihood analysis performed on 100 mock catalogues that contain 1213 galaxies each.   The redshift of each source in the mock catalogues has been randomly assigned following a Gaussian distribution around the best $z_{phot}$. The stellar mass corresponds to a random value taken from another Gaussian distribution, which is centred at the stellar mass given by the random $z_{phot}$.}
\label{fig_mcarlomf} 
\end{center}
\end{figure}

In this section, we investigate in more detail the effect that the uncertainties in the $z_{phot}$ and stellar mass estimates could produce on the galaxy stellar mass function. These uncertainties are the consequence of degeneracies  produced in the SED fitting, and depend on different sources of errors, including the photometric errors, the wavelength coverage, and the limited SED template grid (see e.g. Kitzbichler \& White 2007; Fontanot et al.~2009; Behroozi et al.~2010, for further discussions on this subject).

In particular, we focus here on the analysis of the Eddington bias (Eddington~1913), which can affect the bright end of the galaxy  luminosity or stellar mass function, due to its exponential decline. In simple terms, the $z_{phot}$ and additional stellar mass uncertainties produce a scatter in the derived stellar mass value of each galaxy. But, even if this scatter were symmetric around the best stellar-mass value,  the effect on the steep bright end of the galaxy stellar mass function might not be, as the increment in the number density of sources is relatively more important at higher than lower stellar masses. As a consequence, this effect can produce an artificial flattening of the stellar mass function at the high-mass end.

For simplicity, and to concentrate exclusively on the Eddington bias effect, we performed our analysis considering that the possible $z_{phot}$ for each galaxy follow a Gaussian distribution around the best $z_{phot}$. In each case, the standard deviation $\sigma$ is given by the 68\% confidence values obtained in the SED fitting procedure. We recall that the overall quality of our photometric redshifts is $\rm \sigma[|z_{\rm phot}-z_{\rm spec}|/(1+z_{\rm spec})]=0.05$, after excluding catastrophic outliers (which constitute 3\% of the sample with spectroscopic redshifts). These errors in the $z_{phot}$ directly produce variations in the derived stellar masses. However, in addition, we considered a further error on the stellar mass values, produced by degeneracies in the template parameter space (SED type, age, reddening), at a fixed redshift. Fig.\ref{fig_relerrstm} shows the relative stellar mass error for each of our galaxies that is produced by the degeneracies of the SED template fitting. These errors are $<50\%$ in almost all cases, and $<30\%$ for $\sim$90\% of the most massive galaxies.

We remind that we computed the stellar mass of each galaxy from the best-fitting SED, which is normalised freely between the corresponding $K$ and [4.5] band magnitudes. In this way, the variations in the stellar mass estimates due to degeneracies in parameter space are minimised. Stellar mass estimates which are based on the normalisation of the SED adjusted on UV/optical bands carry much larger uncertainties. In any case, the errors on the stellar mass estimates produced by the uncertainties in the $z_{phot}$ are dominant, and these are governed by the quality of the photometry in all the different bands.

We created a set of mock galaxy catalogues with 1213 galaxies each, where the $z_{phot}$ and stellar masses have been randomly assigned considering two independent Gaussian distributions, as explained above. The Gaussian distribution on the stellar masses has been centred at the stellar mass corresponding to the random $z_{phot}$ in each case, and has a dispersion produced by the template parameter degeneracy at a fixed redshift.

Fig. \ref{fig_mcarlomf} shows the galaxy stellar mass function obtained with the maximum likelihood STY analysis over 100 Monte Carlo realizations, in different redshift bins. The original stellar mass function obtained with the real $z\ge3$ sample is shown for comparison in all cases.

Both at  $3.0 \leq z < 3.5$ and $3.5 \leq z < 4.25$, we see that the galaxy stellar mass function computed on the mock catalogues quite symetrically scatters around the real one at all stellar masses. This shows that the $z_{phot}$/stellar mass  errors do not introduce any significant bias on the determination of the galaxy stellar mass function. This is not surprising, as the combination of large $\alpha$ and $M^\ast$  values indicates that the stellar mass function is characterised by a not so steep decline in the high-mass end. We also note that the scatter of the galaxy stellar mass function produced by the these errors is contained within the $1\sigma$ confidence envelope curves shown in Fig.\ref{fig_mf3z5}.

In the redshift bin $4.25 \leq z < 5.0$, instead, we see that the simulated stellar mass function curves can be flatter than the real one at the high-mass end. As the high-mass end is constrained by less galaxies at such redshifts, the variations in the stellar mass function produced by the $z_{phot}$/stellar mass errors are much more drastic.
This effect is a manifestation of the Eddington bias and, in our case, it appears to affect our $4.25 \leq z < 5.0$ galaxy stellar mass function at $M\gsim 1.5 \times 10^{11} \, \rm M_\odot$.

Correcting the original stellar mass function for the Eddington bias requires some assumption on how the true function can be inferred from the observed one. Following Eddington~(1940; see also Teerikorpi~2004), the true stellar mass function can be approximated by the observed one, convolved by a Gaussian kernel:

\begin{equation}
\Phi^{real}(M) \approx \Phi^{obs.}(M) \times e^{-\frac{1}{2} \sigma^2 \beta^2},
\end{equation}

\noindent where $\sigma$ is the characteristic dispersion observed in the stellar mass. The parameter $\beta$ corresponds to the slope of the stellar mass function bright end, i.e. we can consider $\beta \approx -M/M^\ast$. 

Taking $\sigma=0.3$ (which is roughly the dispersion of $\Delta M/M^\ast$ in our case), we get a correction of  0.03 dex for our $4.25 \leq z < 5.0$  galaxy stellar mass function at $M=1.5 \times 10^{11} \, \rm M_\odot$, and 0.13 dex at $M=3 \times 10^{11} \, \rm M_\odot$.

\subsection{Comparison with other works}
\label{sec_compoth}

\begin{figure}
\begin{center}
\includegraphics[width=70mm]{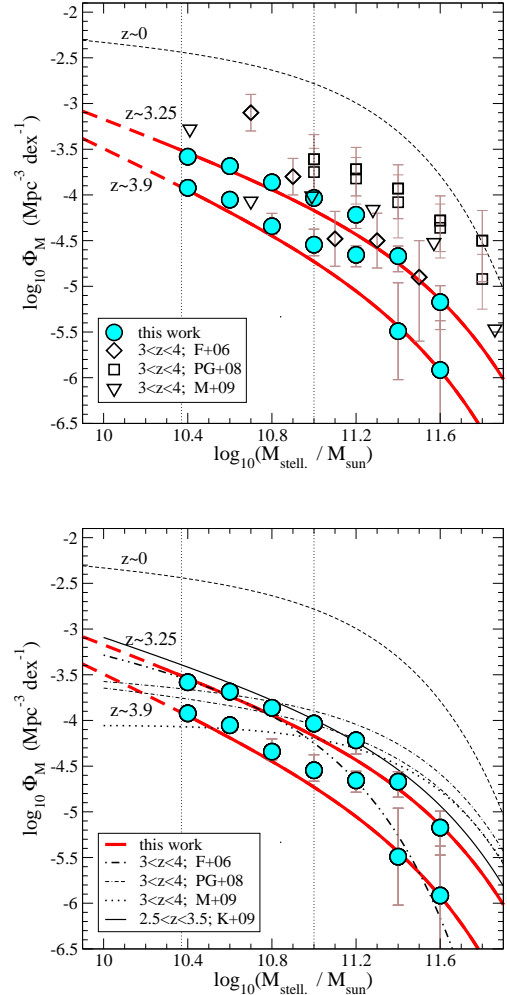}
\caption{Comparison of our stellar mass functions at $3.0 \leq z < 3.5$ and $3.5 \leq z < 4.25$ with those derived by other authors from near or mid-IR-selected galaxy samples [Fontana et al.~(2006; F+06); P\'erez-Gonz\'alez et al.~(2008; PG+08); Kajisawa et al.~(2009; K+09); Marchesini et al.~(2009; M+09) with their model set 8, which corresponds to the same template library, metallicity and reddening law as those we adopt here]. In the upper and lower panels we show the results obtained with the $\rm 1/V_{max}$ method and the maximum likelihood analysis,  respectively.}
\label{fig_compauth}
\end{center}
\end{figure}

It is instructive to compare our stellar mass function with those obtained in previous studies  from both near and mid-IR galaxy-selected samples. This is possible in the redshift range $3<z<4$, where our analysis and other existing studies overlap.

Fontana et al.~(2006) and Kajisawa et al.~(2009) studied the galaxy stellar mass function up to $z\sim3.5-4.0$ within the Great Observatories Origins Deep Survey South and North fields (GOODS-S and GOODS-N), respectively, using $K$-band galaxy selections. Also, P\'erez-Gonz\'alez et al.~(2008) and  Marchesini et al.~(2009) performed similar studies on larger areas ($\sim$ 664 and 511 arcmin$^2$, respectively). The results of these works are compared with our own in Fig. \ref{fig_compauth}. All stellar masses in this figure refer to a Salpeter~(1955) IMF over stellar masses $M=(0.1-100) \, \rm M_\odot$. Stellar masses referring to this IMF are a factor of 1.6 higher than those in the pseudo-Kroupa~(2001) IMF adopted by other authors (such as e.g. Marchesini et al.~2009).

From the two panels of this figure we can see that our estimation of the galaxy stellar mass function at $3 \leq z<4$ (calculated over an area $>$2000 arcmin$^2$) is in very good agreement with those obtained by Fontana et al.~(2006) and Kajisawa et al.~(2009), and in marginal agreement with that of Marchesini et al.~(2009). Note that, for a comparison with Marchesini et al.~(2009), we are considering their stellar mass function derived using the Bruzual \& Charlot~(2007) templates, i.e. the same SED library we adopt in this work. We refer the reader to the Marchesini et al.~(2009) paper for an analysis of the stellar mass function variations produced when using different template libraries, metallicities and reddening laws.

Our stellar mass function is also consistent within the error bars with that of P\'erez-Gonz\'alez et al.~(2008) around a stellar mass $M\sim 10^{11} \, \rm M_\odot$. However, the stellar mass function determined by these authors is significantly discrepant with that of Fontana et al.~(2006) and our own at the high-mass end ($\log_{10}M \gsim 11.4$). Although these differences might be simply due to sample variance effects, we note that our surveyed area is around three times larger than that considered by P\'erez-Gonz\'alez et al.~(2008). Thus, we expect that our determination of the high-mass end of the stellar mass function is more robust.  Moreover, our photometric reshifts have better precision and smaller fraction of catastrophic outliers. This could also explain the discrepancies, because the potential inclusion of a few low-redshift contaminants in their $3<z<4$ sample could be resposible for the enhancement they found at the high-mass end of the stellar mass function.

The comparison of the best parameter values for the maximum likelihood analysis performed with different datasets should be done with care, as the different free parameters are coupled, and different sets of parameter values can produce a similar Schechter function. Keeping this in mind, it is particularly interesting to note the differences found by different authors in the slope $\alpha$ that governs the shape of the stellar mass function at the faint end. P\'erez-Gonz\'alez et al.~(2008) determined that the best value of this slope at $3<z<4$ is  $\alpha \sim 1.2$, i.e. close to the local value, while Fontana et al.~(2006) and Kajisawa et al.~(2009) obtained $\alpha \sim 1.5$ and $1.75$, respectively.   In this work, we find an even higher value for the slope $\alpha=1.86$. As we discussed in Section \S\ref{sec_alphatest}, smaller values such as $\alpha \approx 1.6$ would still be suitable to describe our galaxy stellar mass function, but $\alpha \lsim 1.3$ are clearly rejected from our data. Unfortunately, a more precise constraint on the $\alpha$ value requires a better sampling of the galaxy stellar mass function faint end and, thus, will not be possible until complete, deeper mid-IR surveys become available.

\subsection{The evolution of the stellar mass density with redshift}

\begin{figure}
\begin{center}
\includegraphics[width=75mm]{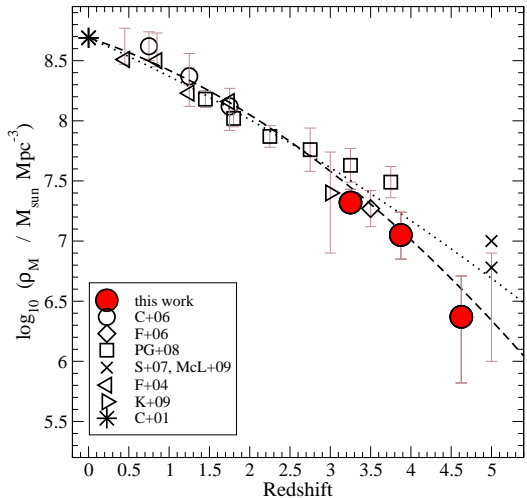}
\caption{Total comoving stellar mass densities versus redshifts. The two curves represent the least-squares fittings to the data points [dashed and dotted lines, respectively for the cases of excluding and including the stellar mass densities estimated at $z\sim5$ (crosses), which are based on optical surveys]. The references in the plot label are the following: Cole et al.~(2001; C+01); Fontana et al. (2004, 2006; F+04, F+06); Caputi et al.~(2006a; C+06); P\'erez-Gonz\'alez et al.~(2008; PG+08); Kajisawa et al.~(2009; K+09); Stark et al.~(2007; S+07); McLure et al.~(2009; McL+09). All stellar mass densities in this diagram correspond to a Salpeter~(1955) IMF over stellar masses $M=(0.1-100) \, \rm M_\odot$.}
\label{fig_stmdens}
\end{center}
\end{figure}

We integrated our resulting stellar mass distribution function $M \,\Phi(M) \, d\log_{10}(M)$, over  stellar masses $\log_{10}(M)=8.0$ to 13.0, in order to obtain estimates of the total comoving stellar mass densities ($\rho_{M}$) at different redshifts. Our results, along with other recent determinations from the literature, are shown in Fig. \ref{fig_stmdens}. 

Our derived comoving stellar mass densities are $\log_{10}(\rho_M)=7.48^{+0.13}_{-0.22}$, $7.05^{+0.11}_{-0.10}$ and $6.37^{+0.14}_{-0.55}$, at redshifts $z \sim 3.25$, 3.9 and 4.6, respectively (where $\rho_M$ is in units of $\rm M_\odot \, Mpc^{-3}$). These values correspond to $\sim$6, 2 and $0.5$\% of the local value. Note that, even at the highest redshifts,  the Eddington bias effect discussed in Section \S\ref{sec_edd} produces a negligible correction, as the contribution of galaxies with stellar masses $M\gsim 1.5 \times 10^{11} \, \rm M_\odot$ to the stellar mass density at $z \sim 4.6$  is only $\sim 4\%$.

Our results are in agreement with those obtained by  Fontana et al.~(2006) at $3<z<4$, Kajisawa et al.~(2009) at $2.5<z<3.5$, and P\'erez-Gonz\'alez et al.~(2008) at $3.0<z<3.5$, within the error bars (note that, for the considered $\alpha$ values, the main contribution to the stellar mass density integral comes from the region $M \lsim M^\ast$, and this is why P\'erez-Gonz\'alez et al. value is consistent with the others in spite of the significant differences in the high-mass end of their stellar mass function).

Instead, our comoving stellar mass density at $4.25 \leq z < 5.0$ is significantly lower than the best estimates of Stark et al.~(2007) and McLure et al.~(2009) at $z\sim5$, based on the study of deep $z$-band-selected galaxy samples  (although we note that our value is still consistent with the lower limit of $\rho_M=10^6 \, \rm  M_\odot \, Mpc^{-3}$ inferred by Stark et al.). We note that, although our galaxy stellar mass function at these redshifts has a similar slope $\alpha$ to that obtained by McLure et al.~(2009), the values of $M^\ast$ and $\Phi^\ast$ are significantly different, explaining the differences observed in the integrated stellar mass densities.

The high uncertainty still existing in the value of the comoving stellar mass density at $z>4$ is  due to different factors. Current mid-IR surveys such as the one analysed here are not sufficiently deep to well sample the stellar mass function at $M \lsim M^\ast$. Instead, optically-selected galaxy samples can reach relatively fainter fluxes and, thus, are more complete for tracing less massive galaxies at high redshifts. However, for most of these less massive galaxies, the rest-frame near-IR emission is unknown, with the result that the stellar mass estimates obtained in optical surveys are very uncertain. Therefore, as we concluded in Section \S\ref{sec_vmaxml}, deeper mid-IR observations will be fundamental to better constrain the stellar mass content locked in galaxies when the Universe was less than 1.0-1.5 Gyr old.

Considering our results along with those compiled from the literature for different redshifts (see Fig. \ref{fig_stmdens}), we can model the evolution of the  stellar mass density from $z\sim5$ to the present day. We perform a least-squares fitting for the data points, assuming a simple functional form for the redshift evolution:

\begin{equation}
\log_{10} \rho_M = az^2 + bz +c,
\end{equation}

\noindent where $a$ and $b$ are free parameters, and $c$ is a constant fixed to the local stellar-mass-density value: $c=8.69$ (Cole et al.~2001). The results of our least-squares fitting  yields: $a=-0.02\pm 0.07$ and $b=-0.30\mp 0.28$, when we consider all the data points; and $a=-0.05\pm 0.09$ and $b=-0.22\mp 0.32$, when we consider only the results of near and mid-IR surveys (i.e. all the stellar mass density estimates except those at $z\sim5$ derived by Stark et al.~2007 and McLure et al.~2009).

\section{IRAC $4.5 \, \rm \mu$\lowercase{m}-selected galaxy candidates at $\lowercase{z}>5$}
\label{sec_zge5}

\begin{figure}
\begin{center}
\includegraphics[width=70mm]{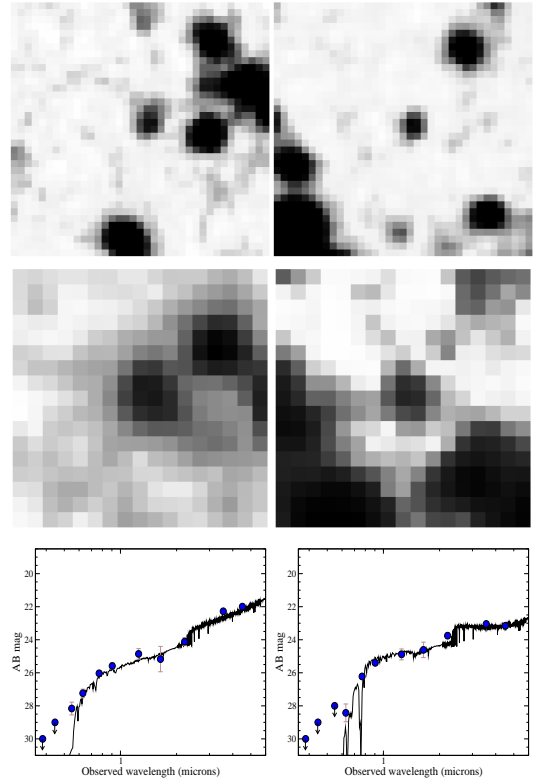}
\caption{Postage stamps for the two candidates in our \ffm sample that are likely genuine $z>5$ galaxies: uds\_ir45\_51741 (left) and uds\_ir45\_96606 (right). The top and middle panels show the $z$-band and  \ffm-band images of these sources, respectively. Each stamp corresponds to an area of $10\times 10$ arscec$^2$. The bottom panels show the correspoding best-fitting SED.}
\label{fig_z5cand}
\end{center}
\end{figure}

As we explained in Section \S\ref{sec_sample}, our final sample contains 10 sources whose best $z_{phot}$ indicate that they should be $z>5$ galaxies. These objects survived all the additional tests we have performed to ensure that we only considered a reliable high-$z$ sample for our analysis. However, it is known that the SEDs of $z>5$ galaxies can be well mimicked by that of other kinds of objects. The main containants are  M, L and T dwarves and intermediate-redshift extremely red galaxies (ERGs; see e.g. Caputi et al.~2004; Simpson et al.~2006). Note that M, L and T dwarves  cannot be easily segregated by making use of a colour-colour diagram such as that shown in Fig. \ref{fig_stargal}.

As discussed by McLure et al.~(2006), an effective discriminant for the $z>5$ galaxy population against lower-$z$ galaxies is the colour-colour criterion $(R-z) \geq 3$ and $(z-J) \leq 1$, but this is not totally effective to separate out M dwarf stars.

Among our ten $z>5$ candidates, one is an $i$-band dropout and another one is a $J$-band dropout within the UDS data depth. These two sources have the typical colours of ERGs, i.e.  $(i-K)>4$ or $(R-K)>5.3$ (Vega), and both are detected in the {\em Spitzer}/MIPS 24 $\rm \mu m$ images with a flux density $S_\nu(24 \, \rm \mu m) \approx 160 \, \mu Jy$. Given these properties, we conclude that these two galaxies are very likely dusty starbursts at intermediate redshifts rather than $z>5$ galaxies.

Other seven out of the ten $z>5$ candidates in our sample have colours $(R-z)< 3$. Six of these are either ERGs defined by $(i-K)>4$ or $(R-K)>5.3$ (Vega), or have an SED that is well fitted with the template of an M dwarf star (from the Rayner, Cushing \& Vacca~2009 library). 

Instead, there is one of these seven sources (id uds\_ir45\_51741, with  $z_{phot}=5.07$ in our catalogue) that is neither an ERG nor any kind of dwarf star. The left-hand panels of Fig. \ref{fig_z5cand} show the $z$-band and \ffm images of this source, as well as its best-fitting SED. Although other objects present in the field are close to this galaxy,  they are at a distance $d>$2 and $d\gsim 4$ arcsec, in the $z$-band and \ffm images, respectively. So,  the aperture photometry for uds\_ir45\_51741 should not significantly be contaminated by the light of neighbour sources.  Therefore, we believe that uds\_ir45\_51741 could be a genuine $z>5$ galaxy in spite of its $(R-z)< 3$ colour. If it were the case,  this would be a rare example of an old and massive galaxy present in the early Universe: the best-fitting SED suggests an age of $\sim$ 1 Gyr and a stellar mass $M \approx 3.8 \times 10^{11} \, M_\odot$.

Finally, one of our 10 $z>5$ candidates (id uds\_ir45\_96606) has $(R-z)=3.03\pm0.56$ and $(z-J)=0.51\pm0.37$. The estimated redshift of this source is $z_{phot}=5.23$. The fitting of its SED with an M dwarf star template can be rejected with more than 3$\sigma$ confidence (with respect to the fitting with a galaxy template). This is a well isolated object in the optical bands, and does not show any sign of blending in the \ffm band (see Fig. \ref{fig_z5cand}). Thus, it is very likely another genuine $z>5$ source. Our SED fitting indicates that this is a $\sim$0.2 Gyr old galaxy with a stellar mass $M \approx 4.6 \times 10^{10} \, M_\odot$.

\begin{figure}
\begin{center}
\includegraphics[width=70mm]{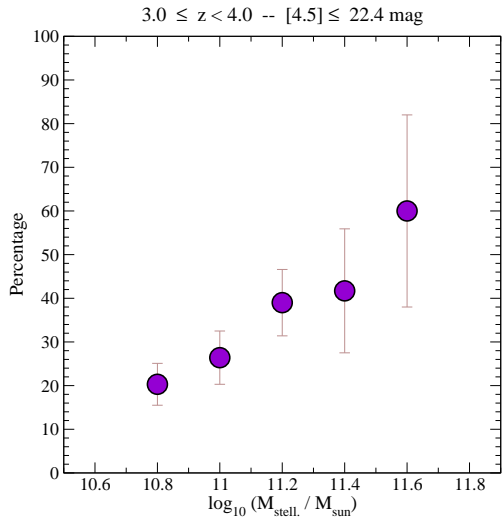}
\caption{The percentage of galaxies with [4.5]$\leq 22.4$ (i.e. within our 80\% completeness limit) at $3\leq z<4$ that have $R>26.6$ and $z>25.6$ AB total magnitudes. These galaxies would be missing in current typical optical surveys, which are brighter than these limits.}
\label{fig_missperc}
\end{center}
\end{figure}

\section{Mid-IR versus optical selection of high-redshift galaxies}
\label{sec_optsel}

The bias introduced by doing a flux-limited \ffm selection of $z\geq 3$ galaxies is quite clear: the most massive galaxies will be included in the sample, while there is a poor constraint on the less massive galaxy populations. Optical selections that map the rest-frame UV light of $z\geq 3$ galaxies, instead, favour the selection of sources with high levels of star formation and/or little dust obscuration, but their biases with respect to a stellar mass selection are not obvious. Our aim in this section is to use our  $z \geq 3$ sample to investigate this problem.

Current optical surveys can reach faint limits, usually $R=27$ or $z=26$ AB magnitudes (measured through 2-arcsec-diameter apertures) over reasonable-size fields, such as the UDS or the COSMOS field. So, within our sample, we searched for galaxies that would be missed even in these deep optical surveys. For a clear comparison, we limited the analysis to our \ffm catalogue with 80\% completeness, i.e. [4.5]$\leq 22.4$, and all redshifts $3 \leq z<4$, to be able to explore a more or less wide stellar mass range.

Fig. \ref{fig_missperc} shows the percentage of our [4.5]$\leq 22.4$ galaxies at $3 \leq z<4$ that have $R>26.6$ and $z>25.6$ AB total magnitudes (roughly equivalent to $R>27$ and $z>26$ 2-arcsec-diameter aperture magnitudes), i.e. that will be missed even by deep optical selections. The total number of \ffm galaxies analysed in each stellar mass bin are 69, 53, 41, 12 and 5 for $\log_{10} M \in [10.7;10.9), [10.9;11.1),[11.1;11.3),[11.3;11.5)$ and  $[11.5;11.7)$, respectively.

Our results show that typical deep optical surveys miss a significant fraction of massive galaxies at $z=3-4$. The percentage of missed galaxies  clearly increases with stellar mass: it is  $(20.3\pm4.8)\%$ for galaxies with $M\sim 6 \times 10^{10} \, M_\odot$, and as high as $(60 \pm 22)\%$ for the rare $M\sim 3-4 \times 10^{11} \, M_\odot$ galaxies.

Note that, although the exact figures depend on the different magnitude cuts chosen for the $R$ and $z$ bands,  our conclusions are still valid even when considering slightly deeper magnitudes. For example, the fraction of galaxies with [4.5]$\leq 22.4$ at $3\leq z<4$ that have $R>27$ and $z>26$ AB total magnitudes range from $(11.6\pm3.9)\%$ to $(29.3 \pm 7.0)\%$ and  $(60 \pm 22)\%$, for stellar masses  $M\sim 6 \times 10^{10}$, $1.6 \times 10^{11}$ and   $M\sim 3-4 \times 10^{11} \, \rm M_\odot$, respectively.

The fact that the fraction of sources missed by deep optical surveys increases with stellar mass is directly related to an increase in internal extinction: the median extinction of our [4.5]$\leq 22.4$ galaxies at $3 \leq z<4$  rises from $A_V=0.80$  for stellar masses $\log_{10} M \in [10.7;10.9)$, to $A_V=2.20$ for the 5 galaxies with $\log_{10} M \in [11.5;11.7)$.

Our results suggest that, at higher redshifts, when galaxies are in general less massive and reddened, deep optical surveys should be less biased in selecting the most massive galaxy populations. At $z\lsim4$, the epoch of obscured star formation and black-hole activity is already on-going, and this is when IR surveys make a key contribution in discovering galaxy populations that are usually not detected otherwise.

\begin{figure}
\begin{center}
\includegraphics[width=85mm]{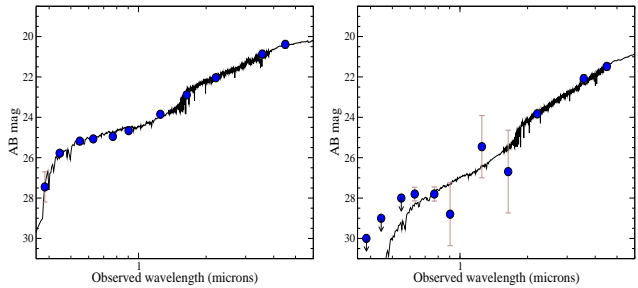}
\caption{The best-fitting SED for two galaxies with $ M > 3.2 \times 10^{11} \, \rm M_\odot$ and $S_\nu(24 \, \rm \mu m) > 300 \, \mu Jy$ in our $3 \leq z<4$ sample. The galaxy corresponding to the left-hand panel would be detected in typical optical surveys, while that of the right-hand panel would be missed.}
\label{fig_massbright24}
\end{center}
\end{figure}

A total of 4 out of the 5 galaxies with $\log_{10} M \in [11.5;11.7)$ in our $3\leq z<4$ sample are detected in  the {\em Spitzer}/MIPS 24 $\rm \mu m$ band with flux densities $S_\nu(24 \, \rm \mu m) > 95 \, \mu Jy$, indicating that these galaxies indeed host  dust-obscured activity. For two of these galaxies, in particular, the 24 $\rm \mu m$ flux densities are 
 quite high ($S_\nu(24 \, \rm \mu m) > 300 \, \mu Jy$), which at these redshifts could suggest the  presence of an obscured active galactic nucleus (AGN; see e.g. Desai et al.~2008). However, we note that, for none of these two galaxies, the UV through near-IR SED shows any obvious sign of an AGN component, i.e. we get good SED fittings using only stellar templates (Fig. \ref{fig_massbright24}).

\section{Summary and Conclusions}
\label{sec_concl}

We have performed a survey of the most massive galaxies present at $z \geq 3$, over an area of 0.6 deg$^2$ of the UKIDSS UDS field. To have the best possible proxy for a stellar mass complete sample, we made our selection in the {\em Spitzer}/IRAC \ffm band, which maps rest-frame near-IR wavelengths at these high redshifts. We followed up our master \ffm catalogue of 50,321 sources in 10 broad bands, from the $U$-band through the IRAC 3.6 $\rm \mu m$ channel. The multi-wavelength follow up has allowed us to model the SEDs of all our galaxies and, with this, obtain redshift estimates and derive stellar masses. Our final sample consists of 1292 galaxies at redshifts $3.0\leq z < 5.23$.

The main goal of our work was the study of the galaxy stellar mass function at $3.0 \leq z<5.0$, particularly the evolution of its high-mass end within this redshift range.  Our deep and homogeneous datasets over a large field are particularly suitable for this purpose. We have found the following:

\begin{itemize}

\item  the comoving number density of the most massive galaxies ($M \gsim 5 \times 10^{10}$ and $M \gsim 1 \times 10^{11} \, \rm M_\odot$ ) declines by a factor $> 10$ within the $\sim$1 Gyr of elapsed time between redshifts $z=3$ and 5. In turn, the number density of  $M \gsim 10^{11} \, \rm M_\odot$ galaxies at $z=3$ is approximately a factor of ten lower than the density $\sim$1 Gyr later at $z=2$, and a factor of $\sim$ 50 lower than in the local Universe  (cf. Caputi et al.~2006a). These results altogether indicate that massive galaxies have assembled at a very fast rate when the age of the Universe was between $\sim$ 1 and 3 Gyr, and this assembly has significantly  slowed down afterwards. 

Our conclusions on the rapid evolution of massive galaxy growth at $2<z<5$ are fully consistent with the intense star formation and black-hole activity characterising this cosmic epoch. At redshifts $z\sim 2-4$, a substantial fraction of massive galaxies are ultra-luminous infrared galaxies (ULIRGs; see e.g. Caputi et al. 2006b; Daddi et al.~2007), and/or host AGN (e.g. Alexander et al.~2005;  Fiore et al.~2008; Yamada et al.~2009). Further investigation of this period appears then to be critical to understand the buildup of the most massive systems and their evolution to the present day.

\item a Schechter function  is still the most suitable form to describe the galaxy stellar mass function at $z\sim4$. However, at $4.25 \leq z <5.0$,  a single power law cannot be discarded with the maximum likelihood analysis. Although this shape degeneracy could be an effect produced by the more restricted sampling that we have around $M  \sim M^\ast$ at such redshifts, it could also be an indication that the shape of the stellar mass function is changing around that epoch. In fact, the transformation of the 
stellar mass function shape is manifested by the progressive steepening of the $\alpha$ value that we discuss below, and the evolution into a single power law will be naturally expected. The possibility of exploring a wider range of stellar masses with deeper mid-IR surveys will be key to probing the transformation of the galaxy stellar mass function shape at high redshifts.

\item the absolute value of the stellar-mass-function faint-end slope $\alpha$ increases with redshift. A similar conclusion has been previously obtained by Fontana et al.~(2006), Kajisawa et al.~(2009) and Marchesini et al.~(2009). In particular, in this work we find that the best values are $1.8< \alpha < 2.1$  at $3 \leq z <5$, and that values as small as $\alpha \lsim 1.3$ can clearly be rejected.

We remark that a steep $\alpha$ value in the Schechter function is a consequence of both the stellar mass function low-mass and high-mass ends. A flat $\alpha \sim 1$, as that characterising the local stellar mass function, is only possible if the stellar mass function has a simple exponential decline at the bright end.  The rise in the $\alpha$ value observed with redshift is not only a consequence of a steepening of the faint end (which, in fact, is poorly constrained by all current surveys), but also indicates {\em a transformation in the high-mass end of the galaxy stellar mass function, with the pure exponential decline no longer being observed at high $z$}.

The steep $\alpha$ values we find at $3 \leq z <5$ are also in line with those obtained for the rest-frame UV luminosity function at $z \gsim 5$. We note, however, that the limits of our mid-IR survey do not allow us to properly constrain the faint end of the galaxy stellar mass function at such high redshifts. This means that a more precise determination of the slope $\alpha$ will only be possible once the galaxy stellar mass function faint end can be better studied with deeper mid-IR surveys.

\item our derived comoving stellar mass densities are $(3.0 \pm 1.2) \times 10^7$ and $(2.36^{+0.89}_{-1.68}) \times 10^6 \, \rm M_\odot Mpc^{-3}$ at redshifts $z\approx 3.25$ and 4.6, respectively, which are around 6 and 0.5\% of the local value. By considering the results of the most recent near- and mid-IR surveys, we found that the redshift evolution of the comoving stellar mass density from $z=0$ to 5 can be modelled as: $\log_{10} \rho_{M} =-(0.05 \pm 0.09) \, z^2 - (0.22\mp 0.32) \, z + 8.69$.

\end{itemize}

Another key result of our work is the absence of massive galaxies at redshifts $z>5$. Within our surveyed area of 0.6 deg$^2$, we find only two quite secure candidates at such high redshifts, and only one with stellar mass $M>10^{11} \, \rm M_\odot$. Instead, optical surveys have discovered a substantial number of intermediate stellar-mass sources at these redshifts. These findings strongly suggest that massive galaxies as a significant population only appear at later times, and that the epoch around redshifts $z\sim3-6$ is critical to understand the formation of the first massive systems.

In addition, we have found that a significant fraction of the most massive galaxies present at $3\leq z<4$ would be missed by optical surveys, even as deep as $R<27$ or $z<26$ magnitudes. This is the consequence of massive galaxies already undergoing the main epoch of obscured star formation and nuclear activity  at these redshifts, which lasts down to redshift $z\sim2$.

Globally, this work has shown that deep mid-IR surveys have a unique role for investigating the first instances of massive galaxy assembly at high redshifts, both through constraining the rest-frame near-IR galaxy light and discovering highly reddened objects. Studying significant areas of the sky  is also key to tracing the most massive galaxies, as they quickly become very rare at high redshifts.

\section*{Acknowledgments}

KIC acknowledges the Leverhulme Trust for funding through the award of an Early Career Fellowship. MC and DF acknowledge the Science \& Technology Facilities Council for funding through the award of an Advanced Fellowship. JSD acknowledges the support of the European Research Council through the award of an Advanced Grant, and the support of the Royal Society through a Wolfson Research Merit Award.  RJM acknowledges the support of the Royal Society through the award of a University Research Fellowship. We are grateful to Peder Norberg for useful discussions. We thank the anonymous referee for a careful review of this paper.

\appendix
\section[]{The impact of sources with degenerate solutions in redshift space on the galaxy stellar mass function at $3\leq\lowercase{z}<5$}

As we explained in Section \ref{sec_zphot}, a total of 244 out of the 1608 galaxies in our initial $z\geq 3$ candidate sample have been discarded on the basis of having degenerate solutions in redshift space, i.e. a redshift $z<3$ could not be rejected within 3$\sigma$ confidence. The fact that these 244 sources are probably not genuine $z\geq 3$ galaxies was supported by the  $z_{\rm phot} - z_{\rm spec}$ diagnostic diagram (see Fig. \ref{fig_zphzsp}), as the percentage of outliers in it would significantly increase if one considered the high-$z$ photometric redshift solutions for these sources.

In this Appendix our aim is to investigate the impact that these discarded sources would have had in our stellar mass function at $3\leq z <5$.

\begin{figure}
\begin{center}
\includegraphics[width=60mm]{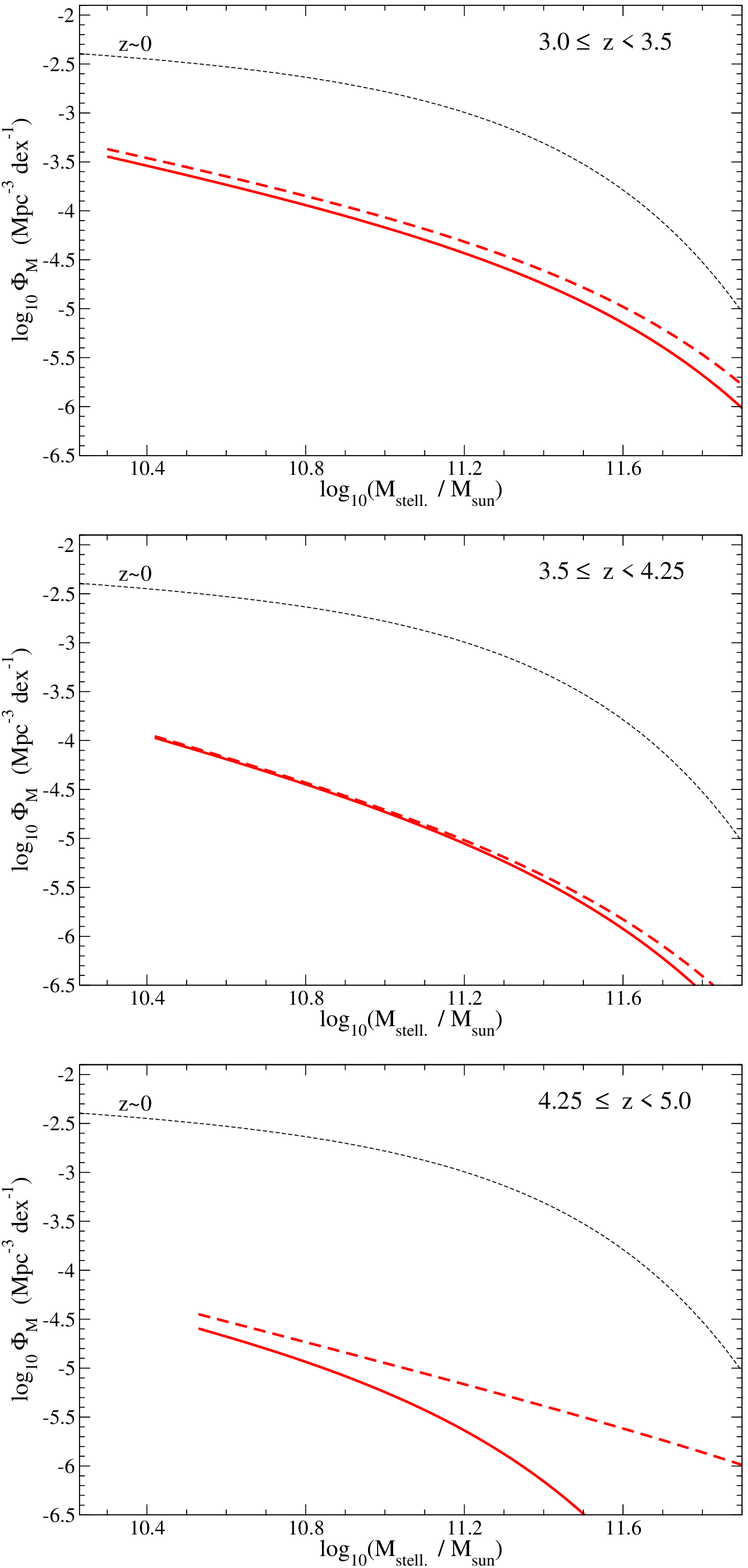}
\caption{Comparison of the galaxy stellar mass function obtained from our secure $z\geq3$ sample (solid lines; see Section \ref{sec_massfun}), with that resulting from incorporating also the extra, unlike $z\geq3$ candidates (dashed lines). These plots illustrate the effect that spurious sources have on the derivation of the stellar mass function.}
\label{fig_appmf}
\end{center}
\end{figure}

Fig. \ref{fig_appmf} shows the stellar mass function that results from the STY maximum likelihood analysis when taking into account the 244 unlike $z\geq 3$ candidates along with our final sample of genuine $z\geq 3$ galaxies.

We see that, for our stellar mass function at  $3.0\leq z <3.5$ and $3.5\leq z <4.25$, the effect of adding these extra sources is almost negligible. The values we obtain for the Schechter function slope and characteristic stellar mass are: $\alpha=1.84^{+0.07}_{-0.05}$, $M^\ast=(3.16^{+1.81}_{-0.80}) \times 10^{11} \, M_\odot$, and $\alpha=2.10^{+0.08}_{-0.07}$, $M^\ast=(2.82^{+1.12}_{-0.60}) \times 10^{11} \, M_\odot$, in the two redshift bins, respectively.  These values are similar to those quoted in Table \ref{table_ml}. This is indicating that the lower redshift contaminants have little impact on our stellar mass function determinations at $3.0\leq z <4.25$, and would evenly populate the galaxy stellar mass function at different stellar masses. 

In the $4.25\leq z <5.0$, instead, the resulting stellar mass function is substantially different to that obtained from the genuine $z\geq 3$ sample. This is because the sample of 244 unlike $z\geq 3$ candidates contains 24 sources with $4.25 \leq z < 5.0$, all of which would have corresponding stellar masses $M> 5\times 10^{10} \, \rm M_\odot$. The addition of these 24 sources triples the population of such massive galaxies at  $4.25 \leq z < 5.0$.

The best Schechter function obtained through the maximum likelihood analysis on the genuine sample and contaminants altogether yields: $\alpha=2.04^{+0.12}_{-0.14}$ and $M^\ast=(2.9^{+3.4}_{-1.7}) \times 10^{12} \, M_\odot$!  This $M^\ast$ value is larger than the stellar mass of any known galaxy at any redshift. Also,  such a stellar mass function would imply  that there are more galaxies with stellar masses $M> 2\times 10^{11} \, \rm M_\odot$ at $4.25 \leq z < 5.0$ than at $3.5 \leq z < 4.25$. These facts strongly indicate that many of the candidates added to the mass function high-mass end are very likely not genuine.

\bsp

\label{lastpage}


\begin{thebibliography}{99}
\bibitem[\protect\citeauthoryear{Alexander}{2005}]{b1} Alexander D.M., Smail I., Bauer F.E., Chapman S.C., Blain A.W., Brandt W.N., Ivison R.J., 2005, Nature, 434, 738
\bibitem[\protect\citeauthoryear{Arnouts}{2007}]{b1}  Arnouts S. et al., 2007, A\&A, 476, 137
\bibitem[\protect\citeauthoryear{Baldry}{2008}]{b1}  Baldry I.K., Glazebrook K., Driver S.P.,  2008, MNRAS, 388, 945
\bibitem[\protect\citeauthoryear{Behroozi}{2010}]{b1} Behroozi P.S., Conroy C., Wechsler R.H.,
2010, ApJ, 717, 379
\bibitem[\protect\citeauthoryear{Bertin}{1996}]{b1}  Bertin E. \& Arnouts S., 1996,
A\&AS, 117, 393
\bibitem[\protect\citeauthoryear{Bolzonella}{2000}]{b1}  Bolzonella M., Miralles J.-M., Pell\'o R., 2000, A\&A, 363, 476
\bibitem[\protect\citeauthoryear{Bolzonella}{2010}]{b1}  Bolzonella M. et al., 2010, A\&A, in press (arXiv:0907.0013)
\bibitem[\protect\citeauthoryear{Bouwens}{2008}]{b1} Bouwens R.J., Illingworth G.D., Franx M., Ford H., 2008, ApJ, 686, 230 
\bibitem[\protect\citeauthoryear{Bruzual}{2003}]{b1} Bruzual G., Charlot S., 2003, MNRAS, 344, 1000
 \bibitem[\protect\citeauthoryear{Bruzual}{2007}]{b1} Bruzual G., 2007, ASPC, 374, 303
\bibitem[\protect\citeauthoryear{Bunker}{2004}]{b1}  Bunker A.J., Stanway E.R., Ellis R.S., McMahon R.G., 2004, MNRAS, 355, 374
\bibitem[\protect\citeauthoryear{Calzetti}{2000}]{b1}  Calzetti D., Armus L., Bohlin R.C., Kinney A.L., Koornneef J., Storchi-Bergmann, T., 2000, ApJ, 533, 682
\bibitem[\protect\citeauthoryear{Caputi}{2004}]{b1}  Caputi K.I., Dunlop J.S., McLure R.J., Roche N.D., 2004, MNRAS, 353, 30 
\bibitem[\protect\citeauthoryear{Caputi}{2005}]{b1}  Caputi K.I., Dunlop J.S., McLure R.J., Roche N.D., 2005, MNRAS, 361, 607
\bibitem[\protect\citeauthoryear{Caputi}{2006a}]{b1}  Caputi K.I., McLure R.J.,
Dunlop J.S., Cirasuolo M., Schael A.M., 2006a, MNRAS, 366, 609
\bibitem[\protect\citeauthoryear{Caputi}{2006b}]{b1} Caputi K.I., Dole H., Lagache, G., McLure, R.J., Dunlop J.S., Puget J.-L., Le Floc'h E., P\'erez-Gonz\'alez P.G., 2006b, A\&A, 454, 143
\bibitem[\protect\citeauthoryear{Caputi}{2007}]{b1}  Caputi K.I. et al., 2007, ApJ, 660, 97 
\bibitem[\protect\citeauthoryear{Cirasuolo}{2007}]{b1}  Cirasuolo M. et al., 2007, MNRAS, 380, 585
\bibitem[\protect\citeauthoryear{Cirasuolo}{2010}]{b1}  Cirasuolo M., McLure R.J., Dunlop J.S., Almaini O., Foucaud S., Simpson C., 2010, MNRAS, 401, 1166  
\bibitem[\protect\citeauthoryear{Cole}{1989}]{b1}  Cole S., Kaiser N., 1989, MNRAS, 237, 1127
\bibitem[\protect\citeauthoryear{Cole}{2001}]{b1}  Cole S. et al., 2001, MNRAS, 326, 255
\bibitem[\protect\citeauthoryear{Daddi}{2005}]{b1}  Daddi E. et al., 2005,
ApJ, 626, 680 
\bibitem[\protect\citeauthoryear{Daddi}{2007}]{b1}  Daddi E. et al., 2007,
ApJ, 670, 156
\bibitem[\protect\citeauthoryear{Desai}{2008}]{b1}  Desai V. et al., 2008, ApJ, 679, 1204
ApJ, 626, 680
\bibitem[\protect\citeauthoryear{Dunlop}{2007}]{b1}  Dunlop J.S., Cirasuolo, M., McLure R.J., 2007, MNRAS, 376, 1054
\bibitem[\protect\citeauthoryear{Dunne}{2009}]{b1} Dunne L. et al., 2009, MNRAS, 394, 3
\bibitem[\protect\citeauthoryear{Eddington}{1913}]{b1} Eddington A.S., 1913, MNRAS, 73, 359
\bibitem[\protect\citeauthoryear{Eddington}{1940}]{b1} Eddington A.S., 1940, MNRAS, 100, 35
\bibitem[\protect\citeauthoryear{Fazio}{2004}]{b1} Fazio G.G. et al., 2004, ApJS, 154, 10
\bibitem[\protect\citeauthoryear{Fiore}{2008}]{b1} Fiore F. et al., ApJ, 672, 94
\bibitem[\protect\citeauthoryear{Fontana}{2004}]{b1}  Fontana A. et al., 2004,
A\&A, 424, 23 
\bibitem[\protect\citeauthoryear{Fontana}{2006}]{b1}  Fontana A. et al., 2006,
A\&A, 459, 745
\bibitem[\protect\citeauthoryear{Fontanot}{2009}]{b1}  Fontanot F., De Lucia G., Monaco P., Somerville R.S., Santini P., 2009, MNRAS, 397, 1776
\bibitem[\protect\citeauthoryear{Furusawa}{2008}]{b1}  Furusawa H. et al., 2008,
ApJS, 176, 1
\bibitem[\protect\citeauthoryear{Hartley}{2010}]{b1} Hartley W. et al., 2010, MNRAS, in press (arXiv:1005.1180)
\bibitem[\protect\citeauthoryear{Ilbert}{2010}]{b1}  Ilbert et al., 2010,
ApJ, 709, 644
\bibitem[\protect\citeauthoryear{Kajisawa}{2009}]{b1} Kajisawa M. et al., 2009, ApJ, 702, 1393
 \bibitem[\protect\citeauthoryear{Kitzbichler}{2007}]{b1} Kitzbichler M.G., White S.D.M., 2007, MNRAS, 376, 2
\bibitem[\protect\citeauthoryear{Kodama}{2007}]{b1}  Kodama T. et al., 2007,
MNRAS, 377, 1717
\bibitem[\protect\citeauthoryear{Kroupa}{2001}]{b1} Kroupa P., 2001, MNRAS, 322, 231
\bibitem[\protect\citeauthoryear{Labb\'e}{2005}]{b1}  Labb\'e et al., 2005,
ApJ, 624, L81
\bibitem[\protect\citeauthoryear{Labb\'e}{2010}]{b1}  Labb\'e et al., 2010,
ApJ, 708, L26
\bibitem[\protect\citeauthoryear{Lawrence}{2007}]{b1}  Lawrence A. et al., 2007,
MNRAS, 379, 1599
\bibitem[\protect\citeauthoryear{Marchesini}{2009}]{b1} Marchesini D., van Dokkum P.G., F\"orster Schreiber N.M., Franx M., Labb\'e I., Wuyts S., 2009, ApJ, 701, 1765
\bibitem[\protect\citeauthoryear{McLure}{2006}]{b1}  McLure R.J. et al., 2006, 
MNRAS, 372, 357
\bibitem[\protect\citeauthoryear{McLure}{2009}]{b1}  McLure R.J., Cirasuolo M., Dunlop J.S., Foucaud S., Almaini O., 2009, MNRAS, 395, 2196
\bibitem[\protect\citeauthoryear{McLure}{2010}]{b1}  McLure R.J., Dunlop J.S., Cirasuolo M., Koekemoer A.M., Sabbi E., Stark D. P., Targett T.A., Ellis R.S., 2010, MNRAS, 403, 960
\bibitem[\protect\citeauthoryear{Mo}{1996}]{b1}  Mo H.J., White S.D.M., 1996,
MNRAS, 282, 347
\bibitem[\protect\citeauthoryear{Oesch}{2010}]{b1} Oesch P. et al., 2010, ApJ, 709, L16
\bibitem[\protect\citeauthoryear{Oke}{1983}]{b1} Oke J.B., Gunn J.E., 1983, ApJ, 266, 713
\bibitem[\protect\citeauthoryear{Ono}{2010}]{b1} Ono Y. et al., 2010, MNRAS, 402, 1580
\bibitem[\protect\citeauthoryear{Papovich}{2006}]{b1}  Papovich et al., 2006, ApJ, 640, 92
\bibitem[\protect\citeauthoryear{Peng}{2010}]{b1} Peng Y. et al., 2010, ApJ, in press (arXiv:1003.4747)
\bibitem[\protect\citeauthoryear{P\'erez-Gonz\'alez}{2008}]{b1}  P\'erez-Gonz\'alez P.G. et al., 2008, ApJ, 675, 234 \bibitem[\protect\citeauthoryear{Pozzetti}{2007}]{b1}  Pozzetti L.. et al., 2007, A\&A, 474, 443
\bibitem[\protect\citeauthoryear{Pozzetti}{2010}]{b1}  Pozzetti L.. et al., 2010, A\&A, in press (arXiv:0907.5416)
\bibitem[\protect\citeauthoryear{Rayner}{2009}]{b1} Rayner J.T.,Cushing M.C., Vacca W.D., 2009, ApJS, 185, 289
\bibitem[\protect\citeauthoryear{Rieke}{2004}]{b1}  Rieke G.H. et al., 2004, ApJS, 154, 25
\bibitem[\protect\citeauthoryear{Rodighiero}{2007}]{b1} Rodighiero G., Cimatti A., Franceschini A., Brusa M., Fritz J., Bolzonella M., 2007, A\&A, 470, 21
\bibitem[\protect\citeauthoryear{Salpeter}{1955}]{b1} Salpeter E.E., 1955, ApJ, 121, 161 
\bibitem[\protect\citeauthoryear{Sandage}{1979}]{b1}  Sandage A., Tammann G.A., Yahil A., 1979, ApJ, 232, 352
\bibitem[\protect\citeauthoryear{Saracco}{2005}]{b1} Saracco P. et al., 2005, MNRAS, 357, L40
\bibitem[\protect\citeauthoryear{Schechter}{1976}]{b1} Schechter P., 1976, ApJ, 203, 297
\bibitem[\protect\citeauthoryear{Schmidt}{1968}]{b1} Schmidt M., 1968, ApJ, 151, 393
\bibitem[\protect\citeauthoryear{Simpson}{2006}]{b1}  Simpson C. et al., 2006, MNRAS, 
373, L21
\bibitem[\protect\citeauthoryear{Stark}{2007}]{b1} Stark D.P., Bunker A.J., Ellis R.S., Eyles, L.P., Lacy M., ApJ, 659, 84
\bibitem[\protect\citeauthoryear{Teerikorpi}{2007}]{b1} Teerikorpi P., 2004, A\&A, 424, 73
\bibitem[\protect\citeauthoryear{Werner}{2004}]{b1}  Werner M.W. et al., 2004, ApJS, 154, 1
\bibitem[\protect\citeauthoryear{Wuyts}{2009}]{b1}  Wuyts S. et al., 2009,
ApJ, 700, 799
\bibitem[\protect\citeauthoryear{Yamada}{2009}]{b1} Yamada T. et al., 2009, ApJ, 699, 1354
\end{thebibliography}
\end{document}